\documentclass[11pt,a4paper]{article} 
\usepackage{jheppub} 
\usepackage{multirow} 

\allowdisplaybreaks[1]

\newcommand\Ref[1] {Ref.\,\cite{#1}}
\newcommand\Refs[2] {Refs.\,\cite{#1} and~\cite{#2}}
\newcommand\Refss[1] {Refs.\,\cite{#1}}

\newcommand\eqn[1] {Eq.\,(\ref{#1})}
\newcommand\eqns[2] {Eqs.\,(\ref{#1}) and~(\ref{#2})}
\newcommand\eqnsnoand[2] {Eqs.\,(\ref{#1}), (\ref{#2})}
\newcommand\eqnss[2] {Eqs.\,(\ref{#1})--(\ref{#2})}

\newcommand\sect[1] {Section~\,{\ref{#1}}}

\newcommand\App[1] {Appendix~\ref{#1}}

\newcommand\tab[1] {Table~\ref{#1}}

\def\beq{\begin{equation}}
\def\eeq{\end{equation}}
\def\bsp#1\esp{\begin{split}#1\end{split}}
\def\bal#1\eal{\begin{align}#1\end{align}}
\newcommand\nt {\notag}

\newcommand\tsig[1] {\sigma^{\mathrm{#1}}} 
\newcommand\dsig[1] {\rd\sigma^{{\rm #1}}} 
\newcommand\dsiga[2] {\rd\sigma^{{\rm #1,A}_{\scriptscriptstyle #2}}} 

\newcommand\la {\langle} 
\newcommand\ra {\rangle} 
\newcommand\bra[3] {\la {\cal M}_{#1}^{#2}#3|} 
\newcommand\ket[3] {|{\cal M}_{#1}^{#2}#3\ra} 
\newcommand{\cM} {{\cal M}} 
\newcommand\M[2] {\ensuremath{|{\cal{M}}_{#1}^{#2}|^2}} 
\newcommand\SME[3] {|{\cal M}_{#1}^{(#2)}{(#3)}|^2} 
 
\newcommand{\PS}[1] {\rd\phi_{#1}} 
\newcommand{\rd} {{\mathrm{d}}} 
\newcommand{\mom}[1] {\{p\}^{#1}} 
\newcommand{\momt}[1] {\{\ti{p}\}^{#1}} 
\newcommand{\momh}[1] {\{\ha{p}\}^{#1}} 
 
\newcommand{\smap}[1] {\stackrel{{\cal S}_{#1}}{\longrightarrow}} 

\newcommand\tzz[2] {z_{#1,#2}}

\newcommand{\calS} {{\cal S}} 

\newcommand{\cC}[2] {{\cal C}_{#1}^{#2}} 
\newcommand{\cS}[2] {{\cal S}_{#1}^{#2}} 
\newcommand{\cSCS}[2] {{\cal C}\kern-2pt{\cal S}_{#1}^{#2}} 

\newcommand{\rC}       {{\mathrm C}}
\newcommand{\rS}       {{\mathrm S}}
\newcommand{\rSCS}     {{\rC}\kern-2pt{\rS}}
 
\newcommand{\IcC}[2] {{\mathrm C}_{#1}^{#2}} 
\newcommand{\IcS}[2] {{\mathrm S}_{#1}^{#2}} 
\newcommand{\IcSCS}[2] {{\mathrm C}\!{\mathrm S}_{#1}^{#2}} 
\newcommand{\bI} {\bom{I}} 

\newcommand\as {\ensuremath{\alpha_{\mathrm{s}}}} 

\newcommand{\CF} {C_{\mathrm{F}}} 
\newcommand{\CA} {C_{\mathrm{A}}} 
\newcommand{\TR} {T_{\mathrm{R}}} 
\newcommand{\Nc} {N_{\mathrm{c}}} 
\newcommand{\Nf} {n_{\mathrm{f}}} 
 
\newcommand{\gam}[1] {\gamma_{#1}} 
\newcommand\gamq[1] {\gamma_q\left(#1\right)}
\newcommand\gamg[1] {\gamma_g\left(#1\right)}
\newcommand\gamf[2] {\gamma_{#1}\left(#2\right)}
\newcommand{\bT} {\bom{T}} 
\newcommand\qb {{\bar q}} 

\newcommand\Oe[1] {\ensuremath{\mathrm O(\ep^{#1})}} 
\newcommand{\cI} {{\cal I}} 
 
\newcommand{\fcut} {\ensuremath{f(y_0,y_{rQ}+y_{sQ}-y_{rs},d'(m,\ep))}}
\newcommand\CUT[1] {\ensuremath{\Theta\left(#1\right)}}

\newcommand\re {{\mathrm e}}
\newcommand{\ep} {\ensuremath{\epsilon}} 
\newcommand{\eps} {\varepsilon} 
\newcommand\al {\ensuremath{\alpha}}

\newcommand{\lam}	{\lambda}
\newcommand{\colx}	{\ensuremath{x}}
\newcommand{\coly}	{\ensuremath{y}}

\def\qqquad{\qquad\quad} 
\def\qqqquad{\qquad\qquad} 

\newcommand\bom[1] {{\mbox{\boldmath $#1$}}}  

\newcommand{\ti}[1] {\tilde{#1}} 
\newcommand{\wti}[1] {\widetilde{\,#1\,}} 
\newcommand{\ha}[1] {\hat{#1}}

\def\AP{Altarelli--Parisi }
\def\MB{Mellin--Barnes }

\newcommand\AB {{\mathrm{ab}}} 
\newcommand\NAB {{\mathrm{nab}}} 

\newcommand\Yt[2]  {Y_{\wti{#1}\wti{#2},Q}}

\title{A subtraction scheme for computing QCD 
jet cross sections at NNLO: integrating 
the doubly unresolved subtraction terms} 
 
\author{G\'abor Somogyi} 
 
\affiliation{PH Department, TH Unit, CERN, CH-1211 Geneva 23, Switzerland} 
 
\emailAdd{Gabor.Somogyi@cern.ch} 

\abstract{We finish the definition of a subtraction scheme for computing 
NNLO corrections to QCD jet cross sections. In particular, we perform the 
integration of the soft-type contributions to the doubly unresolved counterterms 
via the method of \MB representations. With these final ingredients in place, 
the definition of the scheme is complete and the computation of fully differential 
rates for electron-positron annihilation into two and three jets at NNLO accuracy 
becomes feasible. 
}
 
\keywords{QCD, Jets} 
 
\arxivnumber{arXiv:1301.3919} 
 
\preprint{CERN-PH-TH/2013-009}  

\begin{document} 


\maketitle 
\flushbottom


\section{Introduction}
\label{sec:intro}

The high precision of experimental measurements of benchmark processes 
such as jet, heavy quark, vector boson and Higgs boson production at the 
LHC calls for an equally precise theoretical evaluation of relevant 
production rates. In particular, exact perturbative predictions beyond 
next-to-leading order accuracy are desirable and sometimes essential to 
reduce the theoretical uncertainty. Accordingly, in recent years considerable 
effort has beed devoted to computing next-to-next-to-leading order (NNLO) 
corrections to various production and decay rates. Fully differential cross 
sections have been evaluated for vector boson \cite{Melnikov:2006kv,Catani:2009sm}, 
Higgs boson \cite{Anastasiou:2004xq,Catani:2007vq}, diphoton \cite{Catani:2011qz}, 
and Higgs--vector boson \cite{Ferrera:2011bk} production, and for dijet production 
in the fully gluonic channel at leading colour \cite{Ridder:2013mf}. Fully 
differential decay rates for top \cite{Brucherseifer:2013iv} and bottom 
\cite{Brucherseifer:2013cu} quark decay have also been calculated, and 
the computation of the total cross sections for top-antitop production 
\cite{Baernreuther:2012ws,Czakon:2012zr,Czakon:2012pz} and Higgs boson production 
in association with a jet \cite{Boughezal:2013uia} at NNLO is also underway. 
Jet rates and event shapes in electron-positron annihilation to two and three jets 
have also been computed \cite{GehrmannDeRidder:2007jk,GehrmannDeRidder:2008ug,
Weinzierl:2008iv,Weinzierl:2009nz,GehrmannDeRidder:2007bj,GehrmannDeRidder:2007hr,
GehrmannDeRidder:2009dp,Weinzierl:2009ms,Weinzierl:2009yz}.

The construction of a general subtraction scheme to calculate cross sections 
at NNLO accuracy has also received significant attention \cite{Frixione:2004is,
Somogyi:2005xz,GehrmannDeRidder:2005cm,Somogyi:2006da,Somogyi:2006db,
Daleo:2006xa,Somogyi:2008fc,Aglietti:2008fe,Somogyi:2009ri,Bolzoni:2009ye,
Daleo:2009yj,Glover:2010im,Czakon:2010td,Bolzoni:2010bt,Abelof:2011jv,
Gehrmann:2011wi,GehrmannDeRidder:2011aa,GehrmannDeRidder:2012ja,Abelof:2012he}. 
Recall that the the NNLO correction to the cross section for producing $m$ jets 
is a sum of three contributions, the doubly real, real-virtual and doubly virtual 
terms,
\beq
\tsig{NNLO} =
	\int_{m+2}\dsig{RR}_{m+2} J_{m+2}
	+ \int_{m+1}\dsig{RV}_{m+1} J_{m+1}
	+ \int_m\dsig{VV}_m J_m\,,
\label{eq:sigmaNNLO}
\eeq
where each contribution on the right-hand side is separately infrared 
divergent in four space-time dimensions. In \eqn{eq:sigmaNNLO} $J_n$ is
a generic infrared-safe jet measurement function. The essential idea of 
subtraction is that by subtracting and adding back suitable approximate 
cross sections, we can reshuffle divergent pieces among the three terms 
in \eqn{eq:sigmaNNLO}. This reshuffling amounts to writing \eqn{eq:sigmaNNLO} 
as follows,
\beq
\tsig{NNLO} =
	\int_{m+2}\dsig{NNLO}_{m+2}
	+ \int_{m+1}\dsig{NNLO}_{m+1}
	+ \int_m\dsig{NNLO}_m\,,
\label{eq:sigmaNNLOfin}
\eeq
where the three terms
\bal
\dsig{NNLO}_{m+2} &=
	\Big\{\dsig{RR}_{m+2} J_{m+2} - \dsiga{RR}{2}_{m+2} J_{m}
	-\Big[\dsiga{RR}{1}_{m+2} J_{m+1} - \dsiga{RR}{12}_{m+2} J_{m}\Big]
	\Big\}_{\ep=0}\,,
\label{eq:sigmaNNLOm+2}
\\
\dsig{NNLO}_{m+1} &=
	\Big\{\Big[\dsig{RV}_{m+1} + \int_1\dsiga{RR}{1}_{m+2}\Big] J_{m+1} 
	-\Big[\dsiga{RV}{1}_{m+1} + \Big(\int_1\dsiga{RR}{1}_{m+2}\Big)
	\strut^{{\rm A}_{\scriptscriptstyle 1}}
	\Big] J_{m} \Big\}_{\ep=0}\,,
\label{eq:sigmaNNLOm+1}
\eal
and
\beq
\dsig{NNLO}_{m} =
	\Big\{\dsig{VV}_m + \int_2\Big[\dsiga{RR}{2}_{m+2} 
	- \dsiga{RR}{12}_{m+2}\Big] + \int_1\Big[\dsiga{RV}{1}_{m+1} 
	+ \Big(\int_1\dsiga{RR}{1}_{m+2}\Big)
	\strut^{{\rm A}_{\scriptscriptstyle 1}}
	\Big]\Big\}_{\ep=0} J_{m}\,,
\label{eq:sigmaNNLOm}
\eeq
are each integrable in four dimensions by construction \cite{Somogyi:2005xz,
Somogyi:2006cz,Somogyi:2006da,Somogyi:2006db}. The various approximate cross 
sections appearing in \eqnss{eq:sigmaNNLOm+2}{eq:sigmaNNLOm} are as follows.
\begin{itemize}
	\item $\dsiga{RR}{2}_{m+2}$ regulates the doubly unresolved 
	limits of the double real emission cross section, $\dsig{RR}_{m+2}$. 
	\item $\dsiga{RR}{1}_{m+2}$ regulates the singly unresolved 
	limits of the double real emission cross section, $\dsig{RR}_{m+2}$. 
	\item $\dsiga{RR}{12}_{m+2}$ compensates for the double subtraction
	due to the overlap of singly and doubly unresolved limits, i.e.,~it
	regulates the singly unresolved limits of $\dsiga{RR}{2}_{m+2}$ {\em and} 
	the doubly unresolved limits of $\dsiga{RR}{1}_{m+2}$. 
	\item $\dsiga{RV}{1}_{m+1}$ regulates the singly unresolved 
	limits of the real-virtual emission cross section, $\dsig{RV}_{m+1}$. 
	\item $\Big(\int_1\dsiga{RR}{1}_{m+2}\Big)
	\strut^{{\rm A}_{\scriptscriptstyle 1}}$ regulates the singly unresolved 
	limits of the integrated approximate cross section $\int_1 \dsiga{RR}{1}_{m+2}$. 
\end{itemize}
Importantly, just as at NLO accuracy \cite{Frixione:1995ms,Catani:1996vz,
Nagy:1996bz,Somogyi:2006cz,Somogyi:2009ri}, all approximate cross sections 
appearing in \eqnss{eq:sigmaNNLOm+2}{eq:sigmaNNLOm} can be constructed once 
and for all, independently of the process or observable being studied. Indeed, 
the approximate cross sections that contribute to $\dsig{NNLO}_{m+2}$ and 
$\dsig{NNLO}_{m+1}$ in \eqns{eq:sigmaNNLOm+2}{eq:sigmaNNLOm+1} are precisely 
defined in \Refs{Somogyi:2006da}{Somogyi:2006db} respectively. We recall 
that all subtraction terms are constructed using the known universal limit 
formulae for collinear and soft QCD radiation, and their various iterated 
forms \cite{Somogyi:2005xz}. Hence, each approximate cross section is a sum of 
several contributions which may be classified according to the kind of limit 
as collinear- or soft-type.

In order to make the subtraction scheme useful for practical calculations, one 
must also compute the integrals of the approximate cross sections over the phase 
spaces of unresolved partons. The integrated approximate cross section 
$\int_1 \dsiga{RR}{1}_{m+2}$ that appears in \eqn{eq:sigmaNNLOm+1} was evaluated  
in \Ref{Somogyi:2006cz}. The integral of the real-virtual approximate cross 
sections, denoted formally by $\int_1$ in \eqn{eq:sigmaNNLOm}, were computed 
in \Refss{Somogyi:2008fc,Aglietti:2008fe,Bolzoni:2009ye}, while 
$\int_2 \dsiga{RR}{12}_{m+2}$ was evaluated in \Ref{Bolzoni:2010bt}. Finally, 
in \Ref{DelDuca:2013kw}, we calculated all collinear-type contributions to the 
integrated doubly unresolved approximate cross section, 
$\int_2 \dsiga{RR}{2}_{m+2}$. 

In this paper, we complete the definition of the subtraction scheme by 
computing the double soft-type contributions to $\int_2 \dsiga{RR}{2}_{m+2}$. 
We use the method of \MB representations --- developed in this context in 
\Ref{Bolzoni:2009ye} --- to compute the various integrals that appear as 
building blocks for the complete expression. With the results of the present 
paper, the computation of the finite $m$-parton cross section in \eqn{eq:sigmaNNLOm} 
becomes feasible. Since the finiteness of the $m+2$- and $m+1$-parton cross 
sections in \eqns{eq:sigmaNNLOm+2}{eq:sigmaNNLOm+1} was demonstrated in 
\Refs{Somogyi:2006da}{Somogyi:2006db} (for $m=3$ specifically), we are now in 
a position to calculate the fully differential rate for two- and three-jet 
production in electron-positron annihilation at NNLO accuracy within our framework. 
For four or more jets some more work is required, since in \Refs{Bolzoni:2009ye}
{DelDuca:2013kw} (as well as in this paper) a few integrals were computed 
specifically for three-jet kinematics. Having said that, we stress that at 
present the two-loop matrix elements relevant for four or more jet production 
are unknown and these --- rather than the few as yet uncomputed integrals in 
our scheme --- are the essential missing ingredients.

The rest of the paper is organised as follows. In \sect{sec:notation}, we 
set the notation used throughout. The structure of the integrated doubly 
unresolved cross section is recalled next in \sect{sec:double}. It is the 
product (in colour space) of the Born cross section times an appropriate 
insertion operator, which itself is given in terms of so-called flavour-summed  
integrated counterterms. In \sect{sec:intcount} we present these counterterms 
explicitly, recalling their decomposition into non flavour-summed subtraction 
terms, which in turn are defined in terms of integrals of soft currents or 
precisely defined limits of soft currents. These integrals are then expressed 
as linear combinations of a set of basic integrals. Then, in \sect{sec:computing-Srs}  
we discuss our approach to evaluating these basic integrals. Our final results 
are presented in \sect{sec:results}, including analytic expressions for the 
integrated counterterms up to $\Oe{-2}$ and illustrative results for the 
complete insertion operator for two- and three-jet production in electron-positron 
annihilation. Finally, in \sect{sec:conclusions} we draw our conclusions.


\section{Notation}
\label{sec:notation}

%
%

\subsection{Matrix elements and cross sections}

Our notation was spelled out extensively previously
\cite{Bolzoni:2010bt,DelDuca:2013kw} and is recalled here only 
to the extent that we will need in this paper.

The processes we consider are decays of a colourless initial state into any 
number of massless coloured patrons plus any number of additional non-coloured 
final state particles, that are however suppressed in the notation. We use letters 
form the middle of the alphabet, $i$, $j$, $k$, $l$, \ldots, to denote resolved 
final state patrons, while $r$ and $s$ will label the two soft partons. The 
matrix element for a process involving $n$ final state coloured partons will 
be denoted by $\ket{n}{}{}$, using the colour- and spin-state notation of 
\Ref{Catani:1996vz}. The squared matrix element,
\beq
\M{n}{} = \bra{n}{}{}\ket{n}{}{}\,,
\eeq
is understood to be summed over colours and spins, which fixes the normalisation. 
In this paper, we will only need to use tree level matrix elements, which we 
denote $\ket{n}{(0)}{}$. 

Two- and four-parton colour correlated squared tree amplitudes are defined as 
follows
\bal
\M{n,(i,k)}{(0)} &\equiv 
	\bra{n}{(0)}{}\bT_i\cdot \bT_k \ket{n}{(0)}{}\,,
\\
\M{n,(i,k)(j,l)}{(0)} &\equiv 
	\bra{n}{(0)}{}\{\bT_i\cdot \bT_k,\bT_j\cdot \bT_l\} \ket{n}{(0)}{}\,,
\eal
where $\bT_i$ etc.,~is the usual colour-charge operator associated with the 
emission of a gluon from parton $i$. The square of the colour-charge operator, 
$\bT_i^2$, depends only on the flavour of the emitting parton. We emphasise 
this by introducing the notation
\beq
C_{f_i} \equiv \bT_i^2\,,
\eeq
i.e.,~$C_{f_i}$ is the quadratic Casimir operator in the representation 
of parton $i$. Hence, $C_q = \CF=\TR(\Nc^2-1)/\Nc$ and $C_g = \CA=2\TR\Nc$. 
We also use squared colour-charge operators with multiple indices, 
e.g.,~$\bT_{ir}^2 \equiv C_{f_{ir}}$ and $\bT_{irs}^2 \equiv C_{f_{irs}}$. 
In such cases, the multiple index denotes a single parton with flavour obtained 
according to usual flavour summation rules: an odd (even) number of quarks and 
any number of gluons gives a quark (gluon).

Finally, $\dsig{(0)}_n(\mom{})$ denotes the fully differential cross section 
for producing $n$ partons at tree level with momenta $\mom{} \equiv \{p_1,\ldots, 
p_n\}$. Its precise definition reads
\beq
\dsig{(0)}_n(\mom{}) = 
	{\cal N} \sum_{\{n\}} \PS{n}(\mom{};Q) \frac{1}{S_{\{n\}}}
	\SME{n}{0}{\mom{}}\,,
\eeq
where ${\cal N}$ collects factors that are independent of QCD, i.e.,~it 
is the flux factor times the spin average factor for the incoming particles; 
the symbol $\sum_{\{n\}}$ denotes summation over different subprocesses; 
$S_{\{n\}}$ is the Bose symmetry factor for identical particles in the 
final state; and $\PS{n}(\mom{};Q)$ is the phase space associated with 
final state momenta $\mom{}=\{p_1,\ldots,p_n\}$ and total momentum $Q$,
\beq
\PS{n}(p_1,\ldots,p_n;Q) = 
	\prod_{i=1}^n \frac{\rd^d p_i}{(2\pi)^{d-1}} \delta_{+}(p_i^2)\,
	(2\pi)^d \delta^{(d)}\bigg(Q - \sum_{i=1}^n p_i\bigg)\,.
\eeq
In this paper, we will use the cross sections for producing $m$ and $m+2$ 
partons, which we call the Born and doubly real contributions, respectively
\beq
\dsig{B}_m(\mom{}) \equiv \dsig{(0)}_m(\mom{})
\qquad\mbox{and}\qquad
\dsig{RR}_{m+2}(\mom{}) \equiv \dsig{(0)}_{m+2}(\mom{})\,.
\eeq
%

%
%

\subsection{Phase-space factorisation and kinematic variables}

The double soft-type subtraction terms are all defined precisely via the 
double soft momentum mapping of \Ref{Somogyi:2006da}, which maps the original 
set of $m+2$ momenta into a set of $m$ tilded momenta
\beq
\mom{}_{m+2} \smap{rs} \momt{(rs)}_{m}\,.
\label{eq:Srsmap}
\eeq
The explicit form of this mapping can be found in \Ref{Somogyi:2006da}. 
Here we recall only that it leads to an exact factorisation of phase space 
in the following form,
\beq
\PS{m+2}(\mom{}_{m+2};Q) = 
	\PS{m}(\momt{(rs)}_{m};Q) [\rd p_{2;m}^{(rs)}(p_r,p_s;Q)]\,,
\eeq
where the factorised phase-space measure is
\beq
[\rd p_{2;m}^{(rs)}(p_r,p_s;Q)] = 
	\rd (\lam_{rs}^2) (\lam_{rs}^2)^{(m-1)(1-\ep)-1}
	\frac{Q^2}{2\pi} \PS{3}(p_r,p_s,K;Q) 
	\Theta(\lam_{rs})\Theta(1-\lam_{rs})\,,
\label{eq:dPrs2m}
\eeq
with $K^\mu$ a massive momentum such that $K^2 = \lam_{rs}^2 Q^2$.

We will have use for the iterated double soft momentum mapping as well,
\beq
\mom{}_{m+2} 
	\smap{s} \momh{(s)}_{m+1} 
	\smap{\ha{r}} \momt{(\ha{r},s)}_{m}\,,
\label{eq:SrhatSsmap}
\eeq
also defined in \Ref{Somogyi:2006da}. This mapping again leads to the exact 
factorisation of phase space, now in the iterated form
\beq
\PS{m+2}(\mom{}_{m+2};Q) = 
	\PS{m}(\momt{(\ha{r},s)}_{m};Q) 
	[\rd p_{1;m}^{(\ha{r})}(\ha{p}_r;Q)] 
	[\rd p_{1;m+1}^{(s)}(p_s;Q)]\,.
\eeq
The factorised phase-space measures read
\beq
[\rd p^{(t)}_{1;n}(p_t;Q)] = \rd y_{tQ} (1-y_{tQ})^{(n-1)(1-\ep)-1}
	\frac{Q^2}{2\pi} \PS{2}(p_t,K;Q) \Theta(y_{tQ}) \Theta(1-y_{tQ})\,,
\label{eq:dpt1n}
\eeq
where $t=s$ and $n=m+1$ or $t=\ha{r}$ and $n=m$, while $K^\mu$ is a massive 
momentum with $K^2=(1-y_{tQ})Q^2$ and $y_{tQ} \equiv 2p_t\cdot Q/Q^2$. 
The two-parton phase space $\PS{2}(p_t,K;Q)$ can be written as follows in the 
rest frame of $Q$,
\beq
\PS{2}(p_t,K;Q) = 
	\left(\frac{Q^{-2\ep}}{(4\pi)^2} S_\ep\right)
	2^{-1+2\ep} \pi^{\ep} \Gamma(1-\ep)\,
	\rd \eps_t\,\eps_t^{1-2\ep}\delta(\eps_t-y_{tQ})
	\rd \Omega_{d-1}(t)\,,
\label{eq:PS2tn}
\eeq
where $\eps_t = 2 E_t/\sqrt{Q^2}$ is the scaled energy of parton $t$ in 
the $Q$ rest frame, while $\rd \Omega_{d-1}(t)$ is the angular measure in 
$d-1$ dimensions in this frame. In \eqn{eq:PS2tn}, $S_\ep$ is\footnote{$S_\ep
=(4\pi)^\ep \re^{-\ep\gamma_{E}}$ as usually defined in the literature, and 
it is this factor which is conventionally included in the definition of the 
running coupling in the $\overline{\mathrm{MS}}$ renormalisation scheme. Both 
definitions lead to the same expressions in an NLO calculation. At NNLO, the 
two definitions lead to a slightly different bookkeeping of IR and UV poles in 
intermediate steps, but physical cross sections are the same. Our definition 
leads to slightly simpler expressions at NNLO.}
\beq
S_\ep = \frac{(4\pi)^\ep}{\Gamma(1-\ep)}\,.
\eeq

Using the precise definitions of the single and double soft 
momentum mappings \cite{Somogyi:2006da}, it is easy to show that the two 
sets of tilded momenta on the right hand sides of \eqns{eq:Srsmap}
{eq:SrhatSsmap}, $\momt{(rs)}_{m}$ and $\momt{(\ha{r},s)}_{m}$ respectively, 
are related simply by a three-dimensional rotation. This will be a key 
element for the integration of counterterms in \sect{sec:computing-Srs}.

Finally, we recall the definitions of kinematic variables. First, two-particle 
invariants, such as $s_{ik}$, $s_{iQ}$, $s_{\ha{r}Q}$, etc.,~always denote twice 
the dot product of two momenta. A double index enclosed in brackets, as in 
e.g.,~$s_{i(rs)}$, indicates a sum of momenta. Hence
\beq
s_{ik} \equiv 2p_i\cdot p_k\,,
\quad
s_{iQ} \equiv 2p_i\cdot Q\,,
\quad
s_{\ha{r}Q} \equiv 2\ha{p}_r\cdot Q\,,
\quad
s_{i(rs)} \equiv 2p_i\cdot(p_r+p_s)\,,
\quad\mbox{etc.}
\label{eq:sik-def}
\eeq
When these invariants are normalised to the total incoming momentum squared, 
we use the notation $y_{ik} \equiv s_{ik}/Q^2$. The momentum fractions for 
double and triple parton splitting, $\tzz{i}{r}$ and $\tzz{i}{rs}$, are defined 
as follows
\beq
\tzz{i}{r} \equiv \frac{s_{iQ}}{s_{iQ} + s_{rQ}}
\qquad\mbox{and}\qquad
\tzz{i}{rs} \equiv \frac{s_{iQ}}{s_{iQ} + s_{rQ} + s_{sQ}}\,,
\label{eq:zdefs}
\eeq
with $\tzz{r}{i}$ as well $\tzz{r}{is}$ and $\tzz{s}{ir}$ given by obvious 
cyclic permutations. Clearly $\tzz{i}{r}+\tzz{r}{i}=1$ and $\tzz{i}{rs} 
+ \tzz{r}{is} + \tzz{s}{ir}=1$. Furthermore, $\calS_{ik}(r)$ denotes the 
eikonal factor, 
\beq
\calS_{ik}(r) \equiv \frac{2s_{ik}}{s_{ir} s_{kr}}\,.
\label{eq:Sikr-eikonal}
\eeq
Lastly, the integrated counterterms turn out to be functions of the following 
two types of variables
\beq
x_{\wti{i}} \equiv \frac{2\ti{p}_i\cdot Q}{Q^2}
\qquad\mbox{and}\qquad
\Yt{i}{k} \equiv \frac{y_{\wti{i}\wti{k}}}{y_{\wti{i}Q}y_{\wti{k}Q}}\,.
\label{eq:xiYikQdef}
\eeq


\section{The integrated doubly unresolved approximate cross section}
\label{sec:double}

We recall from \Ref{DelDuca:2013kw} that after summing over the flavours 
of unobserved partons, the integrated doubly real approximate cross section 
can be written as
\beq
\int_{2}\dsiga{RR}{2}_{m+2} = 
	\dsig{B}_{m} \otimes \bI^{(0)}_2(\mom{}_{m};\ep)\,.
\label{eq:I1dsigRRA1}
\eeq
Here the insertion operator $\bI^{(0)}_2$ reads
\beq
\bsp
\bI^{(0)}_2(\mom{}_{m};\ep) =
	\left[\frac{\as}{2\pi} S_\ep\left(\frac{\mu^2}{Q^2}\right)^{\ep}\right]^2 
	&\Bigg\{ \sum_{i} \Bigg[
	\IcC{2,i}{(0)} \, \bT_i^2
	+ \sum_{j\ne i} \IcC{2,i j}{(0)}\, \bT_j^2 \Bigg] \bT_i^2
\\ &
	+ \sum_{\substack{j,l \\ j \ne l}}
	\Bigg[
	\IcS{2}{(0),(j,l)}\, \CA\, 
	+ \sum_{i} \IcSCS{2,i}{(0),(j,l)}\, \bT_i^2\, \Bigg] \bT_{j}\bT_{l}
\\ &
	+ \sum_{\substack{i,k \\ i \ne k}}\sum_{\substack{j,l \\ j \ne l}}
	\IcS{2}{(0),(i,k)(j,l)} \{ \bT_{i}\bT_{k},\bT_{j}\bT_{l} \}
	\Bigg\}\,,
\esp
\label{eq:I2}
\eeq
where the sums over $i$, $j$, $k$ and $l$ run over all external partons.
Let us briefly elaborate on the physical content of this formula. As mentioned 
in the Introduction, the approximate cross section $\dsiga{RR}{2}_{m+2}$ is 
actually a sum of terms over the various doubly unresolved kinematical limits 
(triple collinear, double collinear, soft collinear and double soft) and their 
specific iterated forms (which remove multiple subtractions when limits overlap),
\beq 
\bsp 
\dsiga{RR}{2}_{m+2} &= 
	{\cal N} \sum_{\{m+2\}} \PS{m+2}(\mom{};Q) \frac{1}{S_{\{m+2\}}}
\\ &\times
	\sum_{r}\sum_{s\ne r}\Bigg\{ 
	\sum_{i\ne r,s}\Bigg[\frac16\, \cC{irs}{(0,0)}(\mom{})
	+ \sum_{j\ne i,r,s} \frac18\, \cC{ir;js}{(0,0)}(\mom{})
\\ &\qquad\qquad
	+ \frac12\,\Bigg( \cSCS{ir;s}{(0,0)}(\mom{}) 
	- \cC{irs}{}\cSCS{ir;s}{(0,0)}(\mom{}) 
	- \sum_{j\ne i,r,s} \cC{ir;js}{} \cSCS{ir;s}{(0,0)}(\mom{}) \Bigg) 
\\ &\qquad\qquad 
	+ \Bigg( - \cSCS{ir;s}{}\cS{rs}{(0,0)}(\mom{}) 
	- \frac12\, \cC{irs}{}\cS{rs}{(0,0)}(\mom{})
	+ \cC{irs}{}\cSCS{ir;s}{}\cS{rs}{(0,0)}(\mom{}) 
\\ &\qquad\qquad
	+ \sum_{j\ne i,r,s} \frac12\, \cC{ir;js}{}\cS{rs}{(0,0)}(\mom{}) 
	\Bigg)\Bigg] 
	+ \frac12\, \cS{rs}{(0,0)}(\mom{})
	\Bigg\}\,.
\label{eq:dsigRRA2Jmfull} 
\esp 
\eeq 
Integrating the approximate cross section, we obtain the corresponding sum of 
integrated counterterms, 
\beq 
\bsp 
& \int_2 \dsiga{RR}{2}_{m+2} = 
	{\cal N}\sum_{\{m+2\}} \PS{m}(\momt{})\frac{1}{S_{\{m+2\}}} 
	\left[\frac{\as}{2\pi} S_\ep\left(\frac{\mu^2}{Q^2}\right)^{\ep}\right]^2 
\\ &\quad 
	\times \sum_{r}\sum_{s\ne r}\Bigg\{ 
	\sum_{i\ne r,s} \Bigg[ 
	\frac16\,[\IcC{irs}{(0)}]_{f_i f_r f_s} \,(\bT_{irs}^2)^2 
	+ \sum_{j\ne i,r,s} \frac18\,[\IcC{ir;js}{(0)}]_{f_i f_r;f_j f_s} 
	\,\bT_{ir}^2\,\bT_{js}^2  
\\ & \qqqquad\qqquad 
	+ \frac12\, \Bigg( 
	\sum_{\substack{j,l\ne r,s \\ j \ne l}}
	[\IcSCS{ir;s}{(0),(j,l)}]_{f_i f_r}\,\bT^2_{ir}\, \bT_{j} \bT_{l} 
	- [\IcC{irs}{}\!\IcSCS{ir;s}{(0)}]_{f_i f_r}\,(\bT^2_{ir})^2\, 
\\ & \qqqquad\qqqquad 
	- \sum_{j\ne i,r,s} 
	[\IcC{ir;js}{}\!\IcSCS{ir;s}{(0)}]_{f_i f_r}\,\bT^2_{ir}\, \bT^2_{j}\, \Bigg) 
	- \sum_{\substack{j,l\ne r,s \\ j \ne l}}
	[\IcSCS{ir;s}{}\!\IcS{rs}{(0)}]^{(j,l)}\,\bT_{ir}^2\,\bT_{j} \bT_{l} 
\\ & \qqqquad\qqquad 
	+ \bigg([\IcC{irs}{}\!\IcSCS{ir;s}{}\!\IcS{rs}{(0)}]
	- \frac12 [\IcC{irs}{}\!\IcS{rs}{(0)}]_{f_r f_s}\bigg) \big(\bT_{irs}^2\big)^2
	+ \sum_{j\ne i,r,s} \frac12 
	[\IcC{ir;js}{}\!\IcS{rs}{(0)}]\, \bT_{ir}^2\, \bT_{js}^2 \Bigg] 
\\ & \qquad 
	+ \frac12 \sum_{\substack{i,k\ne r,s \\ i \ne k}} 
	\Bigg(\sum_{\substack{j,l\ne r,s \\ j \ne l}}
	[\IcS{rs}{(0)}]^{(i,k)(j,l)}\{\bT_{i} \bT_{k},\bT_{j} \bT_{l}\}
	+ [\IcS{rs}{(0)}]_{f_r f_s}^{(i,k)}\,\CA \bT_{i} \bT_{k} \Bigg) \Bigg\} 
	\otimes  \SME{m}{0}{\momt{}} \,,
\label{eq:IntdsigRRA2} 
\esp 
\eeq 
where we have factored out quadratic Casimir operators to make the 
{\em non flavour-summed} integrated subtraction terms, denoted generically 
as $[X^{(0)}]^{(j,l)\ldots}_{f_i \ldots}$ above, dimensionless in colour 
space. To combine $\int_2 \dsiga{RR}{2}_{m+2}$ in \eqn{eq:IntdsigRRA2} with the 
doubly virtual cross section $\dsig{VV}_{m}$, it must be put in the form of 
an $m$-parton contribution times a factor, which involves rewriting the 
symmetry factor of the $m+2$-parton configuration to the symmetry factor 
of an $m$-parton configuration, i.e.,~summing over unobserved flavours.
Performing this rewriting, we obtain {\em flavour-summed} integrated 
counterterms, denoted $\Big(X^{(0)}\Big)^{(j,l)\ldots}_{f_i\ldots}$, which 
are specific sums of the non flavour-summed integrated subtraction terms, see
\Ref{DelDuca:2013kw} and \sect{sec:intcount} below. Finally, the flavour-summed 
integrated counterterms may be organised according to their structure in colour 
and flavour space into the five terms of \eqn{eq:I2},
\beq
\bsp
\IcC{2,i}{(0)} &=
	\Big(\IcC{irs}{(0)}\Big)_{f_i}
	- \Big( \IcC{irs}{} \IcSCS{ir;s}{(0)} \Big)_{f_i}
	- \Big( \IcC{irs}{} \IcS{rs}{(0)} \Big)_{f_i}
	+ \Big( \IcC{irs}{} \IcSCS{ir;s}{} \IcS{rs}{(0)} \Big)_{f_i}\,,
\\   
\IcC{2,i j}{(0)} &=
	\Big(\IcC{ir;js}{(0)}\Big)_{f_if_j}
	- \Big(\IcC{ir;js}{}\!\IcSCS{ir;s}{(0)}\Big)_{f_if_j}
	+ \Big(\IcC{ir;js}{}\IcS{rs}{(0)}\Big)_{f_if_j}\,,
\\
\IcSCS{2,i}{(0),(j,l)} &=
	\Big(\IcSCS{ir;s}{(0)}\Big)^{(j,l)}_{f_i}
	- \Big(\IcSCS{ir;s}{} \IcS{rs}{(0)}\Big)^{(j,l)}_{f_i}\,,
\\
\IcS{2}{(0),(j,l)} &= 
	\Big(\IcS{rs}{(0)}\Big)^{(j,l)}\,,
\\
\IcS{2}{(0),(i,k)(j,l)} &= 
	\Big(\IcS{rs}{(0)}\Big)^{(i,k)(j,l)}\,.
\esp
\label{eq:cali}
\eeq
Specifically, $\IcC{2,i}{(0)}$ and $\IcC{2,ij}{(0)}$ collect terms that 
come from taking triple and double collinear limits, respectively. They 
depend on the flavour(s) of the hard mother parton(s) in the collinear 
splitting(s) and multiply the Born matrix element with no colour correlations.
All terms coming from taking soft collinear limits are gathered into 
$\IcSCS{2,i}{(0),(j,l)}$ which depends on the flavour of the hard mother 
parton in the collinear splitting and multiplies the two-parton colour correlated 
Born matrix element. Finally, $\IcS{2}{(0),(j,l)}$ and $\IcS{2}{(0),(i,k)(j,l)}$ 
correspond to parts of the double soft limit that multiply the two- and four-parton 
colour correlated Born matrix element, respectively. They are both independent 
of the flavours of hard partons. The structure of each term in colour (flavour) 
space is indicated by upper (lower) indices.

All flavour-summed functions appearing on the right-hand side of \eqn{eq:cali} 
were given in terms of non flavour-summed functions in \Ref{DelDuca:2013kw}, where 
we also computed all those functions in \eqn{eq:cali} which do not involve double 
soft contributions.
In this paper, we compute the remaining flavour-summed double soft functions. 


\section{Integrated double soft-type counterterms}
\label{sec:intcount}

In this section we define explicitly all double soft-type integrated 
counterterms that appear on the right-hand side of \eqn{eq:cali}. In each 
case, we first present the flavour decomposition of the flavour-summed 
counterterms. Then, we give the proper definition of the non flavour-summed 
functions in terms of integrals of soft currents or precisely defined limits 
of soft currents. Finally, we compute the results in terms of a set of basic 
integrals.

\paragraph{Double soft:} 
The flavour decomposition of the double soft counterterms reads
\beq
\bsp
\Big(\IcS{rs}{(0)}\Big)^{(i,k)(j,l)} &=
	\frac{1}{2} [ \IcS{rs}{(0)} ]_{gg}^{(i,k)(j,l)}\,,
\\
\Big(\IcS{rs}{(0)}\Big)^{(j,l)} &=
	\frac12\, [ \IcS{rs}{(0)} ]^{(j,l)}_{gg}
	+ \Nf\, [ \IcS{rs}{(0)} ]^{(j,l)}_{\qb q}\,,
\esp
\label{eq:Srs0}
\eeq
where $\Nf$ is the number of light quark flavours. The functions appearing 
on the right-hand side of \eqn{eq:Srs0} are defined as follows
\bal
[ \IcS{rs}{(0)} ]^{(i,k)(j,l)}_{gg} &=
	\left( \frac{(4\pi)^2}{S_\ep}Q^{2\ep}\right)^2
	\int_2 [\rd p_{2;m}^{(rs)}(p_r,p_s;Q)]\,
	\frac{1}8 \calS_{ik}(r)\calS_{jl}(s)
\nt\\&\times
	\fcut\,,
\label{eq:defSikjl}
\\
[ \IcS{rs}{(0)} ]^{(i,k)}_{gg} &=
	- \left( \frac{(4\pi)^2}{S_\ep}Q^{2\ep}\right)^2
	\int_2 [\rd p_{2;m}^{(rs)}(p_r,p_s;Q)]\,
	\frac{1}4 \calS_{ik}(r,s)
\nt\\&\times
	\fcut\,,
\label{eq:defSikgg}
\\
[ \IcS{rs}{(0)} ]^{(i,k)}_{\qb q} &=
	\left( \frac{(4\pi)^2}{S_\ep}Q^{2\ep}\right)^2
	\int_2 [\rd p_{2;m}^{(rs)}(p_r,p_s;Q)]
	\frac{1}{s_{rs}^2}\, \frac{\TR}{\CA}\,
\nt\\&\times
	\Bigg( \frac{s_{ir}s_{ks}+s_{kr}s_{is}-s_{ik}s_{rs}}{s_{i(rs)}s_{k(rs)}}
	-2\frac{s_{ir}s_{is}}{s_{i(rs)}^2}\Bigg)
	\fcut\,.
\label{eq:defSikqq}
\eal
The non-abelian part of the double soft gluon current, $\calS_{ik}(r,s)$ 
in \eqn{eq:defSikgg}, reads \cite{Catani:1999ss}
\beq
\calS_{ik}(r, s) = 
	\calS_{ik}^{(\rm s.o.)}(r, s) + 4 \frac{s_{ir} s_{ks} 
	+ s_{is} s_{kr}}{s_{i(rs)} s_{k(rs)}} \left[\frac{1-\ep}{s_{rs}^2} 
	- \frac18 \calS_{ik}^{(\rm s.o.)}(r, s)\right] - \frac{4}{s_{rs}} 
	\calS_{ik}(rs)\ ,
\label{eq:softggnab}
\eeq
where
\beq
\calS_{ik}^{(\rm s.o.)}(r, s) = 
	\calS_{ik}(s)\left(\calS_{is}(r) + \calS_{ks}(r) - \calS_{ik}(r)\right)
\label{eq:softggnabso}
\eeq
is the form of this function in the strongly-ordered limit (either $E_r \ll E_s$ 
or $E_s \ll E_r$ as the expression is symmetric in $r$ and $s$ when summed over 
$i$ and $k$) and $\calS_{ik}(rs)$ is given by
\beq
\calS_{ik}(rs) = 
	\frac{2 s_{ik}}{s_{i(rs)} s_{k(rs)}}\,.
\label{eq:Sikrs}
\eeq
Finally, the function $\fcut$ appearing in \eqnss{eq:defSikjl}{eq:defSikqq} 
represents a small but convenient modification of the subtraction terms as 
compared to their original definitions in \Ref{Somogyi:2006da}. Its precise 
role and form are explained in \App{app:modifiedA2soft}, and in the following, 
we will include this factor without further comment.

In this paper, we do not discuss the case when $i$, $k$, $j$ and $l$ in 
\eqn{eq:defSikjl} are all distinct, this requiring at least four jets at NNLO. 
For the specific cases of  two and three hard partons, the integrated double 
soft counterterms are computed in \App{app:RR_A2_Srs} and we find
\bal
[\IcS{rs}{(0)}]^{(i,k)(j,k)}_{gg} &= 
	\cI_{2\cS{}{},1}(\Yt{i}{k},\Yt{i}{j},\Yt{j}{k},\ep;y_0,d'_0)\,,
\label{eq:ISrs0ikjkMIexp}
\\
[\IcS{rs}{(0)}]^{(i,k)(i,k)}_{gg} &= 
	\cI_{2\cS{}{},2}(\Yt{i}{k},\ep;y_0,d'_0)\,,
\label{eq:ISrs0ikikMIexp}
\\
[\IcS{rs}{(0)}]_{gg}^{(i,k)} &= 
	\cI_{2\cS{}{},2}(\Yt{i}{k},\ep;y_0,d'_0)
	-\cI_{2\cS{}{},3}(\Yt{i}{k},\ep;y_0,d'_0)
\nt\\&
	-2\cI_{2\cS{}{},4}(\Yt{i}{k},\ep;y_0,d'_0)
	+\cI_{2\cS{}{},5}(\Yt{i}{k},\ep;y_0,d'_0)
\nt\\&
	+4\cI_{2\cS{}{},6}(\Yt{i}{k},\ep;y_0,d'_0)
	+2(1-\ep)\cI_{2\cS{}{},7}(\Yt{i}{k},\ep;y_0,d'_0)
\nt\\&
	-(1-\ep)\cI_{2\cS{}{},8}(\ep;y_0,d'_0)\,,
\label{eq:ISrs0ggikMIexp}
\\
[\IcS{rs}{(0)}]_{q\qb}^{(i,k)} &= 
	\frac{\TR}{\CA}\Big[
	-\cI_{2\cS{}{},6}(\Yt{i}{k},\ep;y_0,d'_0)
	-2\cI_{2\cS{}{},7}(\Yt{i}{k},\ep;y_0,d'_0)
\nt\\&\qquad\qquad
	+2\cI_{2\cS{}{},9}(\ep;y_0,d'_0)
	\Big]\,.
\label{eq:ISrs0qqikMIexp}
\eal
The integrals $\cI_{2\cS{}{},n}$ for $n=1,\ldots,9$ are defined in 
\eqnss{eq:IS1}{eq:IS9}.

\paragraph{Triple collinear -- double soft:}
The flavour decomposition of the triple collinear -- double soft counterterm 
reads
\beq
\Big( \IcC{irs}{} \IcS{rs}{(0)} \Big)_f =
	\frac{1}{2} [ \IcC{irs}{} \IcS{rs}{(0)} ]_{fgg}
	+ \Nf\, [ \IcC{irs}{} \IcS{rs}{(0)} ]_{f\qb q}\,.
\label{eq:CirsSrs0f}
\eeq
We further decompose the triple collinear -- double soft gluon counterterm 
as a sum of abelian and non-abelian pieces,
\beq
[ \IcC{irs}{} \IcS{rs}{(0)} ]_{fgg} =
	[ \IcC{irs}{} \IcS{rs}{(0)} ]_{gg}^{(\AB)}
	+ [ \IcC{irs}{} \IcS{rs}{(0)} ]_{fgg}^{(\NAB)}\,,
\eeq
following the decomposition of the triple \AP splitting kernels in a similar 
fashion \cite{Catani:1999ss}. Then we have the following explicit definitions 
for the non flavour-summed functions
\bal
[ \IcC{irs}{} \IcS{rs}{(0)} ]_{gg}^{(\AB)} &=
	\left( \frac{(4\pi)^2}{S_\ep}Q^{2\ep}\right)^2
	\int_2 [\rd p_{2;m}^{(rs)}(p_r,p_s;Q)]\,
	\frac{4\tzz{i}{rs}^2}{s_{ir}s_{is}\tzz{r}{is}\tzz{s}{ir}}
\nt\\&\times
	\fcut\,,
\label{eq:defC3Sab}
\\
[ \IcC{irs}{} \IcS{rs}{(0)} ]_{fgg}^{(\NAB)} &=
	\left( \frac{(4\pi)^2}{S_\ep}Q^{2\ep}\right)^2
	\int_2 [\rd p_{2;m}^{(rs)}(p_r,p_s;Q)]\,
	\frac{\CA}{\bT_{irs}^2}\,
\nt\\&\times
	\bigg[
	\frac{(1 - \ep)}{s_{i(rs)} s_{rs}}
	\frac{(s_{ir} \tzz{s}{ir} - s_{is} \tzz{r}{is})^2}
     {s_{i(rs)} s_{rs} (\tzz{r}{is} + \tzz{s}{ir})^2}
	- \frac{\tzz{i}{rs}}{s_{i(rs)} s_{rs}}
	\left(\frac{4}{\tzz{r}{is}+\tzz{s}{ir}} - \frac1{\tzz{r}{is}} \right)
\nt\\&\qquad
	- \frac{1}{s_{i(rs)} s_{ir}} \frac{2\tzz{i}{rs}^2}
	{\tzz{r}{is} (\tzz{r}{is} + \tzz{s}{ir})}
	- \frac{\tzz{i}{rs}^2}{s_{i(rs)} s_{is}} \frac{1}{\tzz{r}{is} (\tzz{r}{is} 
	+ \tzz{s}{ir})} 
\nt\\&\qquad
	+ \frac{\tzz{i}{rs}}{s_{ir} s_{rs}}
	\left(\frac{1}{\tzz{s}{ir}} + \frac{1}{\tzz{r}{is} + \tzz{s}{ir}}\right)
	+ (r \leftrightarrow s)
	\bigg]\, 
\nt\\&\times
	\fcut\,,
\label{eq:defC3Snab}
\\
[ \IcC{irs}{} \IcS{rs}{(0)} ]_{f\qb q} &= 
	\left( \frac{(4\pi)^2}{S_\ep}Q^{2\ep}\right)^2
	\int_2 [\rd p_{2;m}^{(rs)}(p_r,p_s;Q)] 
	\frac{2}{s_{i(rs)}\,s_{rs}}\, \frac{\TR}{\bT_{irs}^2}
\nt\\&\times
	\Bigg( \frac{\tzz{i}{rs}}{\tzz{r}{is} + \tzz{s}{ir}}
	-\frac{(s_{ir}\tzz{s}{ir} - s_{is}\tzz{r}{is})^2}
	{s_{i(rs)}s_{rs}(\tzz{r}{is} + \tzz{s}{ir})^2} \Bigg)\, 
\nt\\&\times
	\fcut\,.
\label{eq:defC3Sqq}
\eal
The integrated triple collinear -- double soft counterterms are computed 
in \App{app:RR_A2_CirsSrs}. In terms of the basic integrals introduced in 
\eqnss{eq:IS10}{eq:IS18}, we find
\bal
[\IcC{irs}{}\IcS{rs}{(0)}]_{gg}^{(\AB)} &= 
	4 \cI_{2\cS{}{},10}(\ep;y_0,d'_0)\,,
\label{eq:ICirsSrs0ggABMIexp}
\\
[\IcC{irs}{}\IcS{rs}{(0)}]_{fgg}^{(\NAB)} &= 
	\frac{\CA}{\bT_{irs}^2}\Big[
	4(1-\ep) \cI_{2\cS{}{},11}(\ep;y_0,d'_0)
	-4(1-\ep) \cI_{2\cS{}{},12}(\ep;y_0,d'_0)
\nt\\&\qquad\qquad
	-8 \cI_{2\cS{}{},13}(\ep;y_0,d'_0)
	+2 \cI_{2\cS{}{},14}(\ep;y_0,d'_0)
	-4 \cI_{2\cS{}{},15}(\ep;y_0,d'_0)
\nt\\&\qquad\qquad
	-2 \cI_{2\cS{}{},16}(\ep;y_0,d'_0)
	+2 \cI_{2\cS{}{},17}(\ep;y_0,d'_0)
	+2 \cI_{2\cS{}{},18}(\ep;y_0,d'_0)\Big]\,,	
\label{eq:ICirsSrs0ggNABMIexp}
\\
[\IcC{irs}{}\IcS{rs}{(0)}]_{fq\qb} &= 
	\frac{\TR}{\bT_{irs}^2}\Big[
	-4 \cI_{2\cS{}{},11}(\ep;y_0,d'_0)
	+4 \cI_{2\cS{}{},12}(\ep;y_0,d'_0)
	+2 \cI_{2\cS{}{},13}(\ep;y_0,d'_0)\Big]\,.
\label{eq:ICirsSrs0qqMIexp}
\eal
%

\paragraph{Double collinear -- double soft:}
The flavour decomposition of the double collinear -- double soft counterterm 
is trivial,
\beq
\Big(\IcC{ir;js}{}\IcS{rs}{(0)}\Big)_{f_1f_2} =
	\frac{1}{2} [ \IcC{ir;js}{}\IcS{rs}{(0)} ]\,,
\label{eq:CirjsSrs0ff}
\eeq
and the counterterm is in fact independent of the flavours $f_1$ and $f_2$. 
The precise definition of $[ \IcC{ir;js}{}\IcS{rs}{(0)} ]$ on the right-hand 
side of \eqn{eq:CirjsSrs0ff} reads
\beq
\bsp
[ \IcC{ir;js}{}\IcS{rs}{(0)} ] &=
	\left( \frac{(4\pi)^2}{S_\ep}Q^{2\ep}\right)^2
	\int_2 [\rd p_{2;m}^{(rs)}(p_r,p_s;Q)]\,
	\frac{1}{s_{ir}}\frac{2\tzz{i}{r}}{\tzz{r}{i}}\,
	\frac{1}{s_{js}}\frac{2\tzz{j}{s}}{\tzz{s}{j}}
\\&\times
	\fcut\,.
\esp
\label{eq:defC2S}
\eeq
This integral is computed in \App{app:RR_A2_CirjsSrs}.  The result is
\beq
[\IcC{ir;js}{}\IcS{rs}{(0)}] = 
	4 \cI_{2\cS{}{},19}(\Yt{i}{j},\ep;y_0,d'_0)\,,
\label{eq:ICirjsSrs0MIexp}
\eeq
where $\cI_{2\cS{}{},19}$ is defined in \eqn{eq:IS19}.

\paragraph{Soft collinear -- double soft:}
The flavour decomposition of the soft collinear -- double soft 
counterterm is also trivial,
\beq
\Big(\IcSCS{ir;s}{} \IcS{rs}{(0)}\Big)^{(j,l)}_f =
	[ \IcSCS{ir;s}{}\IcS{rs}{(0)} ]^{(j,l)}\,,
\label{eq:CSirsSrs0f}
\eeq
and the counterterm is in fact independent of the flavour $f$. The non 
flavour-summed counterterm on the right-hand side of \eqn{eq:CSirsSrs0f} 
is defined precisely as follows 
\beq
\bsp
[ \IcSCS{ir;s}{}\IcS{rs}{(0)} ]^{(j,l)} &=
	- \left( \frac{(4\pi)^2}{S_\ep}Q^{2\ep}\right)^2
	\int_2 [\rd p_{2;m}^{(rs)}(p_r,p_s;Q)]\,
	\,\frac{1}{s_{ir}}\frac{2\tzz{i}{r}}{\tzz{r}{i}}
	\frac{1}2 \calS_{jl}(s)
\\&\times
	\fcut\,.
\label{eq:defCSSjl}
\esp
\eeq
We remind the reader that above $j\ne l$, but $(ir)$ may be equal to $j$ or $l$. 
The cases when $j,l \ne (ir)$ and when e.g.,~$j=(ir)$ lead to different integrals, 
as discussed in \App{app:RR_A2_CSirsSrs}. In terms of the basic integrals 
introduced in \eqns{eq:IS20}{eq:IS21}, the result is
\beq
[\IcSCS{ir;s}{}\IcS{rs}{(0)}]^{(j,l)} = 
	-2 \cI_{2\cS{}{},20}(\Yt{j}{l},\ep;y_0,d'_0)\,,
\label{eq:ICSirsSrs0jlMIexp}
\eeq
if $(ir)$ is distinct form both $j$ and $l$, while for e.g.,~$j=(ir)$, we find
\beq
[\IcSCS{ir;s}{}\IcS{rs}{(0)}]^{(i,l)} = 
	-2 \cI_{2\cS{}{},21}(x_{\wti{i}},\Yt{i}{l},\ep;y_0,d'_0)\,.
\label{eq:ICSirsSrs0irlMIexp}
\eeq
%

\paragraph{Triple collinear -- soft collinear -- double soft:}
Finally, the flavour decomposition of the triple collinear -- soft collinear 
-- double soft counterterm is again trivial,
\beq
\Big( \IcC{irs}{} \IcSCS{ir;s}{} \IcS{rs}{(0)} \Big)_f =
	[ \IcC{irs}{} \IcSCS{ir;s}{} \IcS{rs}{(0)} ]\,,
\label{eq:CirsCSirsSrs0ff}
\eeq
and the counterterm is in fact independent of the flavour $f$. The function 
$[ \IcC{irs}{} \IcSCS{ir;s}{} \IcS{rs}{(0)} ]$ in \eqn{eq:CirsCSirsSrs0ff} 
is defined as
\beq
\bsp
[ \IcC{irs}{} \IcSCS{ir;s}{} \IcS{rs}{(0)} ] &=
	\left( \frac{(4\pi)^2}{S_\ep}Q^{2\ep}\right)^2
	\int_2 [\rd p_{2;m}^{(rs)}(p_r,p_s;Q)]\,
	\frac{4 \tzz{i}{rs}(\tzz{i}{rs} + \tzz{r}{is})}
	{s_{ir}\,s_{(ir)s}\,\tzz{r}{is}\,\tzz{s}{ir}}\, 
\\&\times
	\fcut\,.
\esp
\label{eq:defCCSSgg}
\eeq
This integral is computed in \App{app:RR_A2_CirsCSirsSrs}. We find
\beq
[\IcC{irs}{}\IcSCS{ir;s}{}\IcS{rs}{(0)}] = 
	4 \cI_{2\cS{}{},22}(x_{\wti{i}},\ep;y_0,d'_0)
	+ 4 \cI_{2\cS{}{},23}(x_{\wti{i}},\ep;y_0,d'_0)\,,
\label{eq:ICirsCSirsSrs0MIexp}
\eeq
with $\cI_{2\cS{}{},22}$ and $\cI_{2\cS{}{},23}$ defined in 
\eqns{eq:IS22}{eq:IS23}.
%


\section{Computing the double soft integrals}
\label{sec:computing-Srs}

%
%

\subsection{Simplifying the integrals}

Consider a generic double soft-type master integral,
\beq
\cI_{2\cS{}{},n} = \left(\frac{(4\pi)^2}{S_\ep} Q^{2\ep}\right)^2
	\int_2[\rd p^{(rs)}_{2;m}]\,
	F_n(\{s_{jl},z_j\};\ep)
	\fcut\,,
\label{eq:ISi-general}
\eeq
where $F_n(\{s_{jl},z_j\};\ep)$ is the integrand, which will depend on 
two-particle invariants $s_{jl}$ and possibly also momentum fractions $z_j$. 
(We use $z_j$ generically to denote both the two- and three-parton momentum 
fractions defined in \eqn{eq:zdefs}.) A straightforward treatment of 
\eqn{eq:ISi-general} proceeds to first express the integrand in terms of 
independent (tilded) momenta. Then, choosing some particular Lorentz frame, 
the integral is written in terms of e.g.,~the angles and energies of momenta 
$p_r^\mu$ and $p_s^\mu$ in this frame. The result obtained however turns out 
to be completely unwieldy. In particular, the extraction of $\ep$ poles via 
sector decomposition is not possible, because one typically finds complicated 
singularities inside the domain of integration. It also turns out that such a 
parametrisation is not a useful starting point for deriving \MB representations.

A much more manageable form is obtained, however, by observing that the value of 
the integral in \eqn{eq:ISi-general} is unchanged if we replace the double soft 
unresolved phase-space measure $[\rd p^{(rs)}_{2;m}]$ by the iterated single soft phase-space measure, e.g.,~$[\rd p_{1;m}^{(\ha{r})}] [\rd p_{1;m+1}^{(s)}]$.

To understand why this may be the case, note that by using the precise 
definitions of the single and double soft momentum mappings \cite{Somogyi:2006da}, 
it is easy to show that the two sets of tilded momenta $\momt{(rs)}_{m}$ and 
$\momt{(\ha{r},s)}_{m}$, where
\beq
\mom{}_{m+2} \smap{rs} \momt{(rs)}_{m}
\qquad\mbox{and}\qquad
\mom{}_{m+2} \smap{s} \momh{(s)}_{m+1} \smap{\ha{r}} \momt{(\ha{r},s)}_{m}\,,
\eeq
are related simply by a three-dimensional rotation. This implies that the 
dot products of momenta in each set are the same for both sets. But the 
integrated counterterms depend only on these dot products (as opposed to the 
overall orientation of the tilded momenta), hence we are free to use the more 
convenient iterated phase-space mapping when computing the integrals,
\beq
\bsp
&\int_2[\rd p^{(rs)}_{2;m}]\,
	F_n(\{s_{jl},z_j\};\ep)
	\fcut 
\\ &\qquad=
\int_2[\rd p_{1;m}^{(\ha{r})}] [\rd p_{1;m+1}^{(s)}]\,
	F_n(\{s_{jl},z_j\};\ep)
	\fcut\,.
\label{eq:dp2-dp1dp1}
\esp
\eeq
By using the iterated form of the phase-space measure, the integrations over 
the variables of $\ha{p}_r^\mu$ and $p_s^\mu$ may be performed sequentially. 

%
%

\subsection{Computing the integrals via \MB representations}

Following the discussion of the previous section, we write the generic 
master integral in \eqn{eq:ISi-general} as
\beq
\cI_{2\cS{}{},n} = \left(\frac{(4\pi)^2}{S_\ep} Q^{2\ep}\right)^2
	\int_2[\rd p^{(\ha{r})}_{1;m}] [\rd p^{(s)}_{1;m+1}]\,
	F_n(\{s_{jl},z_j\}) \fcut\,.
\eeq
Since the dimension of $F_n$ is $[F_n] = (Q^2)^{-2}$, we have
\beq
F_n(\{s_{jl},z_j\}) = \frac{1}{(Q^2)^2} F_n(\{y_{jl},\tzz{j}{kl}\})\,,
\eeq
and hence, using \eqnsnoand{eq:dpt1n}{eq:PS2tn} and the exact definition 
of $\fcut$ in \eqn{eq:fcut-1}, we find
\beq
\bsp
\cI_{2\cS{}{},n} &= 2^{-4+4\ep} \pi^{-2+2\ep} \Gamma^2(1-\ep)
	\int_{0}^{1} \rd y_{\ha{r}Q}\, y_{\ha{r}Q}^{1-2\ep} 
	(1-y_{\ha{r}Q})^{d'_0-2+\ep} 
	\int_{0}^{1} \rd y_{sQ}\, y_{sQ}^{1-2\ep} (1-y_{sQ})^{d'_0-1}
\\&\times
	\CUT{y_0-y_{\ha{r}Q}-y_{sQ}+y_{\ha{r}Q}y_{sQ}}
	\rd\Omega_{d-1}(\ha{r}) 	\rd\Omega_{d-1}(s)  F_n(\{y_{jl},z_j\})\,.
\esp
\label{eq:Srs-generic}
\eeq
In the above form, \eqn{eq:Srs-generic} is directly suitable for treatment 
by \MB techniques. In particular, the angular integrations over the directions 
of $\ha{p}_r^\mu$ and $p_s^\mu$ may be preformed sequentially using the results 
of \Ref{Somogyi:2011ir}. Hence, it becomes essentially straightforward to derive 
a \MB representation for the integral:%
\footnote{Note that contrary to the phase space integrals considered here, 
\MB representations for loop Feynman integrals can be constructed automatically,  
e.g.,~with the {\tt AMBRE.m} package \cite{Gluza:2007rt}.}
\begin{enumerate}
\item Substitute the specific form of $F_n$.

\item Perform the angular integration over $\rd\Omega_{d-1}(s)$, using the 
results of \Ref{Somogyi:2011ir}. The momenta which are to be kept fixed during 
this integration are the intermediate, hatted momenta of \eqn{eq:SrhatSsmap}. 
Thus, at this stage, all invariants need to be expressed in terms of these:
\beq
y_{ik} = (1-y_{sQ}) y_{\ha{i}\ha{k}}\,,
\quad
y_{is} = y_{\ha{i}s}\,,
\quad
y_{iQ} = (1-y_{sQ})y_{\ha{i}Q} + y_{\ha{i}s}\,,
\qquad i,k\ne s\,.
\label{eq:plain2hat}
\eeq
The result after this angular integration depends on (scaled) dot products of 
hatted momenta and perhaps $Q$, i.e.,~$y_{\ha{i}\ha{k}}$, $y_{\ha{i}\ha{r}}$, 
$y_{\ha{i}Q}$, etc.  

\item Repeat the previous step for the angular integrals over 
$\rd\Omega_{d-1}(\ha{r})$. The independent momenta with regard to the
$\ha{p}_r^\mu$ integration are now the final set of tilded momenta in 
\eqn{eq:SrhatSsmap}, so we must express all invariants in terms of these:
\beq
y_{\ha{i}\ha{k}} = (1-y_{\ha{r}Q}) y_{\ti{i}\ti{k}}\,,
\quad
y_{\ha{i}\ha{r}} = y_{\ti{i}\ha{r}}\,,
\quad
y_{\ha{i}Q} = (1-y_{\ha{r}Q})y_{\ti{i}Q} + y_{\ti{i}\ha{r}}\,,
\qquad i,k\ne r\,.
\label{eq:hat2tilde}
\eeq
After this integration, the expression depends only on invariants involving 
tilded momenta, as well as $y_{\ha{r}Q}$ and $y_{sQ}$, which are however 
integration variables themselves.

\item Finally, perform the integrations over $y_{\ha{r}Q}$ and $y_{sQ}$ to 
obtain the full \MB representation. If $y_0=1$, the argument of the $\Theta$ 
function in \eqn{eq:Srs-generic} factorises and these last two integrations 
are automatically in the form of a beta function integral. For $0<y_0<1$, 
additional \MB integrations must be introduced to bring them to this form.
\end{enumerate}

With this procedure, we obtain a multi-dimensional \MB integral representation 
of \eqn{eq:Srs-generic}. It is usually not possible to evaluate these integrals 
analytically. However, for practical purposes, we are actually interested in the 
$\ep$-expansion of $\cI_{2\cS{}{},n}$, rather than the all-orders result. 
Importantly, this expansion may be performed in an algorithmic way, e.g.,~with 
the {\tt MB.m} \cite{Czakon:2005rk} or {\tt MBresolve.m} \cite{Smirnov:2009up} 
packages. Then we obtain the expansion coefficients as finite \MB integrals. 
We are able to evaluate all integrals that contribute to the $1/\ep^{4}$ and 
$1/\ep^{3}$ poles analytically,\footnote{In these calculations, the {\tt 
barnesroutines.m} package of D.~Kosower is useful for applying the first and 
second Barnes lemmas on \MB integrals automatically.} while the rest of the 
expansion coefficients can be computed by direct numerical integration of 
the \MB representation.

Finally, we call attention to the following technical detail. We find that it 
is best to fix the specific value of $d'_0$ (see \eqn{eq:dv0def}) {\em before} 
the $\ep$-expansion of the integrals. Therefore, in the following, we set
$d'_0 = 3-3\ep$, i.e.,~$D'_0=3$ and $d'_1=-3$. This is not an essential restriction 
and recomputing the expansions for different values of $D'_0$ and $d'_1$ is in 
principle straightforward.


\section{Results}
\label{sec:results}

%
%

\subsection{Analytic expressions to $\Oe{-2}$}

We have obtained analytic expressions for the integrated counterterms 
up to $\Oe{-2}$ accuracy. The results below are quoted for $d'_0=3-3\ep$, 
i.e.,~$D'_0=3$ and $d'_1=-3$ (see \eqn{eq:dv0def}), and generic $y_0\in (0,1]$. 
Before presenting the actual formulae, we call attention to the following 
features. Up to this pole order:
\begin{itemize}
\item All kinematic dependence enters through logarithms of the variables 
$x_i$ and/or $Y_{ik,Q}$ defined in \eqn{eq:xiYikQdef}.

\item Dependence on the cut parameter $y_0$ always enters in the same 
functional form, as follows,
\beq
\ln y_0 - 2 y_0 +\frac{y_0^2}{2}\,.
\eeq
We observe that this is simply $\Sigma(y_0,2)$ (recall that throughout we 
use $D'_0=3$), with the $\Sigma(z,N)$ function of 
\Refs{Somogyi:2008fc}{Bolzoni:2010bt},
\beq
\Sigma(z,N) = \ln z - \sum_{k=1}^{N} \frac{1-(1-z)^k}{k}\,.
\eeq

\item Additional constants that enter are always recognised to be slight 
generalisations of $\gamma_q/C_q$ and $\gamma_g/C_g$ (since our integrated 
counterterms are dimensionless in colour space), where $\gamma_g = \frac{3}{2}\CF$ 
and $\gamma_g=\frac{11}{6}\CA-\frac{2}{3}\TR\Nf$ \cite{Kunszt:1992tn}. In order 
to emphasise this connection, we define the following two functions of $\Nf$ 
\cite{DelDuca:2013kw},
\beq
\gamq{\Nf} = \frac{3}{2}
\qquad\mbox{and}\qquad
\gamg{\Nf} = \frac{11}{6} - \frac{2}{3}\frac{\TR}{\CA} \Nf\,.
\eeq
The formal $\Nf$ dependence of $\gamq{\Nf}$ is introduced in order to make 
possible a flavour-independent notation in the following.
\end{itemize}
We find
\begin{enumerate}
\item Double soft:
\bal
\Big(\IcS{rs}{(0)}\Big)^{(i,k)(j,k)}(Y_{ik,Q},Y_{ij,Q},Y_{jk,Q}) &=
	{1 \over 4}\bigg[{1 \over \ep^4}
	- {1 \over \ep^3}\bigg(\ln Y_{ik,Q} + \ln Y_{jk,Q} 
	+ 4 \Sigma(y_0,2)\bigg)\bigg]
\nt\\&
	+ \Oe{-2}\,,
\label{eq:ISrs0ikjkAN}
\intertext{note the lack of $Y_{ij,Q}$ dependence to this order,}
\Big(\IcS{rs}{(0)}\Big)^{(i,k)(i,k)}(Y_{ik,Q}) &=
	{1 \over 4}\bigg[{1 \over \ep^4}
	- {1 \over \ep^3}\bigg(2\ln Y_{ik,Q} 
	+ 4 \Sigma(y_0,2)\bigg)\bigg]
	+ \Oe{-2}\,,
\label{eq:ISrs0ikikAN}
\eal
and
\beq
\Big(\IcS{rs}{(0)}\Big)^{(i,k)}(Y_{ik,Q}) =
	- {1 \over 4}\bigg[{1 \over \ep^4}
	- {1 \over \ep^3}\bigg(2\ln Y_{ik,Q} + 4 \Sigma(y_0,2) 
	- \gamg{\Nf}\bigg)\bigg]
	+ \Oe{-2}\,.
\label{eq:ISrs0ikAN}
\eeq

\item Triple collinear -- double soft:
\beq
\Big(\IcC{irs}{}\IcS{rs}{(0)}\Big)_{f_i} =
	\frac{1}{2}\left[\left(1 + \frac{\CA}{2C_{f_i}}\right)
	\left(\frac{1}{\ep^4} - \frac{4}{\ep^3} \Sigma(y_0,2)\right)
	+\frac{1}{\ep^3}\frac{\CA}{2C_{f_i}} \gamf{g}{\Nf}\right] 
	+ \Oe{-2}\,.
\label{eq:ICirsSrs0fAN}
\eeq
Note that this counterterm does not depend on kinematics.

\item Double collinear -- double soft:
\beq
\Big(\IcC{ir;js}{}\IcS{rs}{(0)}\Big)_{f_i f_j}(Y_{ij,Q}) =
	\frac{1}{2}\left[
	{1 \over \ep^4} - {4 \over \ep^3} \Sigma(y_0,2)\right]
	+ \Oe{-2}\,.
\label{eq:ICirjsSrs0ffAN}
\eeq
Note the lack of $Y_{ij,Q}$ dependence to this order.

\item Soft collinear -- double soft: 
\bal
\Big(\IcSCS{ir;s}{}\IcS{rs}{(0)}\Big)_{f_i}^{(j,l)}(Y_{jl,Q}) &=
	-\bigg[{1 \over \ep^4} 
	- {1 \over \ep^3}\bigg(\ln Y_{jl,Q} + 4 \Sigma(y_0,2)\bigg)\Bigg]
	+ \Oe{-2}
\label{eq:ICSirsSrs0fjl}
\intertext{for $j,l \ne i$, and}
\Big(\IcSCS{ir;s}{}\IcS{rs}{(0)}\Big)_{f_i}^{(i,l)}(x_i,Y_{il,Q}) &=
	-{1 \over 6}\bigg[{5 \over \ep^4} 
	- {1 \over \ep^3}\bigg(\ln x_i + 6\ln Y_{il,Q} 
	+ 19 \Sigma(y_0,2)\bigg)\Bigg]
	+ \Oe{-2}
\label{eq:ICSirsSrs0firl}
\eal
for e.g.,~$j=i$.

\item Triple collinear -- soft collinear -- double soft:
\beq
\Big(\IcC{irs}{}\IcSCS{ir;s}{}\IcS{rs}{(0)}\Big)_{f_i}(x_i) =
	\frac{1}{3}\left[{2 \over \ep^4} 
	- {1 \over \ep^3} \bigg(\ln x_i 
	+ 7\, \Sigma(y_0,2)\bigg)\right]
	+ \Oe{-2}\,.
\label{eq:ICirsCSirsSrs0fAN}
\eeq
\end{enumerate}

Substituting \eqnss{eq:ISrs0ikjkAN}{eq:ICirsCSirsSrs0fAN} and the corresponding 
results from \Ref{DelDuca:2013kw} --- recalled here in \App{app:collICTs} for the 
convenience of the reader --- into \eqn{eq:cali}, we obtain explicit expressions 
for the kinematics-dependent functions entering the insertion operator. We note 
that since in this paper we set $D'_0=3$, for consistency we must also use the 
results of \Ref{DelDuca:2013kw} with $D'_0=3$. The results do not depend on $d_0$ 
and $\al_0$ up to this order.\footnote{$d_0$ and $\al_0$ are  
parameters of the collinear-type counterterms that correspond to the $d'_0$ and 
$y_0$ used in this paper. In particular, $\al_0\in (0,1]$, if smaller than one, 
restricts subtractions to near collinear regions in phase space. See 
\Ref{DelDuca:2013kw} for further details.} 
Starting with $\IcC{2,i}{(0)}$, we find
\beq
\IcC{2,i}{(0)}(x_i) =
	-{1 \over 2\ep^3}\bigg(1 + {\CA \over C_{f_i}}\bigg)
	\bigg[2\ln x_i - 2 \Sigma(y_0,2) 
	- \gamf{f_i}{\Nf {\CF \over C_{f_i}}}\bigg]
	+ \Oe{-2}\,.
\label{eq:IC20f}
\eeq
For $\IcC{2,i j}{(0)}$, we obtain
\beq
\IcC{2,i j}{(0)}(x_i, x_j) =
	{1 \over 2\ep^3}\bigg[
	\Big(2\ln x_i - \gamf{f_i}{\Nf}\Big)
	-\Big(2\ln x_j - \gamf{f_j}{\Nf}\Big)\bigg]
	+ \Oe{-2}\,.
\label{eq:IC20ff}
\eeq
However, $\IcC{2,i j}{(0)}$ enters the insertion operator $\bI_2^{(0)}$ 
summed over its flavour indices, thus we are free to symmetrise in $i$ and 
$j$. In particular, setting
\bal
\IcC{2,(i j)}{(0)}(x_i, x_j) \equiv 
\frac{1}{2}\Big(\IcC{2,i j}{(0)}(x_i, x_j)+\IcC{2,j i}{(0)}(x_j, x_i)\Big)\,,
\eal
(we use the usual notation of round brackets around indices to denote 
symmetrisation) we see that up to this order in the $\ep$ expansion, 
$\IcC{2,(i j)}{(0)}$ simply vanishes,
\beq
\IcC{2,(i j)}{(0)}(x_i, x_j) = \Oe{-2}\,.
\label{eq:IC20ffsymm}
\eeq
For $\IcSCS{2,i}{(0),(j,l)}$ we find (note the vanishing of $Y_{jl,Q}$ 
dependence to this order)
\bal
\IcSCS{2,i}{(0),(j,l)}(x_i, Y_{jl,Q}) &=
	{1 \over \ep^3}\bigg(
	2\ln x_i - 2\Sigma(y_0,2) - \gamf{f_i}{\Nf}\bigg) 
	+ \Oe{-2}
\label{eq:ICS20ijl}
\intertext{when $j$ and $l$ are distinct form $i$, and}
\IcSCS{2,i}{(0),(i,l)}(x_i, Y_{il,Q}) &=
	{3 \over 4\ep^3}\bigg(
	2\ln x_i  - 2\Sigma(y_0,2) - \gamf{f_i}{\Nf}\bigg)
	+ \Oe{-2}
\label{eq:ICS20iil}
\eal
for e.g.,~$j=i$.
The two-parton colour-correlated soft function, $\IcS{2}{(0),(j,l)}$, is
\beq
\IcS{2}{(0),(j,l)}(Y_{jl,Q}) = 
	- {1 \over 4}\bigg[{1 \over \ep^4}
	- {1 \over \ep^3}\bigg(2\ln Y_{jl,Q} + 4 \Sigma(y_0,2)	
	- \gamg{\Nf}\bigg)\bigg]
	+ \Oe{-2}\,.
\label{eq:IS20jl}
\eeq
Last, we consider the four-parton colour-connected soft function, 
$\IcS{2}{(0)(i,k)(j,l)}$. We find (note that $Y_{ij,Q}$ dependence 
is absent to this order)
\beq
\IcS{2}{(0),(i,k)(j,k)}(Y_{ik,Q},Y_{ij,Q},Y_{jk,Q}) =
	{1 \over 4}\bigg[{1 \over \ep^4}
	- {1 \over \ep^3}\bigg(\ln Y_{ik,Q} + \ln Y_{jk,Q} + 4 \Sigma(y_0,2)\bigg)\bigg]
	+ \Oe{-2}\,,
\label{eq:IS20ikjk}
\eeq
if three indices are distinct, while
\beq
\IcS{2}{(0),(i,k)(i,k)}(Y_{ik,Q}) =
	{1 \over 4}\bigg[{1 \over \ep^4}
	- {1 \over \ep^3}\bigg(2\ln Y_{ik,Q} + 4 \Sigma(y_0,2)\bigg)\bigg]
	+ \Oe{-2}\,,
\label{eq:IS20ikik}
\eeq
if only two indices are different.

%
%

\subsection{Insertion operator for two- and three-jet production}

Using the results of \Ref{DelDuca:2013kw} and the present paper, we 
can assemble the complete insertion operator $\bI_2^{(0)}$ relevant 
for two- and three-jet production. 

Let us consider first the process $e^+e^- \to 2$ jets. The corresponding
squared matrix element at tree level is $|\cM^{(0)}_2(1_q,2_\qb)|^2$,
i.e.,~the quark carries label 1 and the antiquark label 2. Both the colour
algebra and kinematics are trivial. Colour conservation implies
\beq
\bT_1\,\bT_2 = -\CF
\eeq
and $C_{f_1} = C_{f_2} = \CF$, while momentum conservation gives 
\beq
x_1 = x_2 = Y_{12,Q} = y_{12} = 1\,.
\eeq
Then the insertion operator, \eqn{eq:I2}, is a scalar in colour space and 
reads
\beq
\bsp
\bI^{(0)}_{2} =
	\left[\frac{\as}{2\pi}S_\ep \left(\frac{\mu^2}{Q^2}\right)^\ep\right]^2
	&\bigg[
	2 \CF^2\left(\rC^{(0)}_{2,q} + \rC^{(0)}_{2,q q}
	- 2 \rSCS^{(0),(1,2)}_{2,q} + 4 \rS^{(0),(1,2)(1,2)}_{2}\right)
\\&
	- 2 \CF \CA \rS^{(0),(1,2)}_{2}
\bigg]\,,
\esp
\label{eq:I22jet}
\eeq
with all arguments being equal to one. Using \eqnss{eq:IC20f}{eq:IS20ikik},  
we find
\beq
\bI_2^{(0)}(p_1,p_2;\ep) = 
	\left[\frac{\as}{2\pi}S_\ep \left(\frac{\mu^2}{Q^2}\right)^\ep\right]^2
	\CF^2\left\{
	\frac{4 + x}{2\ep^4} + 
	\frac{72 + 29 x - 4 y \Nf}{12\ep^3}
	+ \Oe{-2} \right\} \,,
\eeq
where we used the notation \cite{Nagy:1997md}
\beq
\colx = \frac{\CA}{\CF}\,,
\qquad 
\coly = \frac{\TR}{\CF}\,.
\eeq
Notice that up to this order, $\bI_2^{(0)}$ is independent of $y_0$.

The remaining expansion coefficients are computed numerically. We present these 
in \tab{tab:I22jet} for the quantity $\bI_2^{(0)}(p_1,p_2;\ep)/\CF^2$. In this 
calculation, we have set $d_0=d'_0=3-3\ep$ and $\al_0=y_0=1$, however colour 
factors and the number of light flavours, $\Nf$, were kept symbolic. Hence the 
actual value of the expansion coefficient at a given order in $\ep$ is given by 
the scalar product of the vector of numbers in the appropriate column of  
\tab{tab:I22jet} with the vector of coefficients forming the first column.
\begin{table}[t]
\begin{center}
\renewcommand{\arraystretch}{2}
\begin{tabular}{|c|c|c|c|}
\hline\hline
coeff. & $1/\ep^{2}$ &  $1/\ep^{1}$ &  finite
\\
\hline\hline
\multirow{1}{*}{$1$} 
 & -7.416 $\pm$ 0.011
 & -81.383 $\pm$ 0.052
 & -236.572 $\pm$ 0.328
\\ 
\hline 
\multirow{1}{*}{$\frac{\CA}{\CF}$} 
 & -13.028 $\pm$ 0.003
 & -66.274 $\pm$ 0.025
 & -281.208 $\pm$ 0.188
\\ 
\hline 
\multirow{1}{*}{$\frac{\Nf\TR}{\CF}$} 
 & 1.000 $\pm$ 0.001
 & 9.984 $\pm$ 0.010
 & 57.535 $\pm$ 0.075
\\ 
\hline\hline
\end{tabular}
\caption{
\label{tab:I22jet}
Coefficients of the Laurent expansion of 
$\displaystyle\frac{\bI_2^{(0)}(p_1,p_2;\ep)}{\CF^2}$ for $e^+e^- \to 2$ 
jet production. We used $d_0=d'_0=3-3\ep$ and $\al_0=y_0=1$.}
\end{center}
\end{table}

Turning to the process $e^+e^- \to 3$ jets, the corresponding squared 
matrix element at tree level is $|\cM^{(0)}_2(1_q,2_\qb,3_g)|^2$,
i.e.,~the quark carries label 1, the antiquark label 2 and the gluon
carries label 3. The colour algebra is still trivial and using colour 
conservation we find 
\beq
\bT_1\,\bT_2 = \frac{\CA - 2\CF}{2}\,,\qquad
\bT_1\,\bT_3 = \bT_2\,\bT_3 = -\frac{\CA}{2}\,,
\eeq
and $C_{f_1} = C_{f_2} = \CF$, while $C_{f_3} = \CA$. The insertion operator, 
\eqn{eq:I2}, is again a scalar in colour space and we have
\beq
\bsp
\bI^{(0)}_{2} =
	\left[\frac{\as}{2\pi}S_\ep \left(\frac{\mu^2}{Q^2}\right)^\ep\right]^2
	&\bigg\{
	\CF^2 \Big(\rC^{(0)}_{2,q}(x_1) 
	+ \rC^{(0)}_{2,q}(x_2)\Big)
	+ \CA^2 \rC^{(0)}_{2,g}(x_3)
\\&
	+ \CF^2 \Big(\rC^{(0)}_{2,qq}(x_1,x_2) 
	+ \rC^{(0)}_{2,qq}(x_2,x_1)\Big)
\\&
	+ \CF \CA \Big(\rC^{(0)}_{2,qg}(x_1,x_3) 
	+ \rC^{(0)}_{2,gq}(x_3,x_1)
\\&\qquad\qquad
	+ \rC^{(0)}_{2,qg}(x_2,x_3) 
	+ \rC^{(0)}_{2,gq}(x_3,x_2)\Big)
\\&
	+ (\CA - 2 \CF) \Big[\CA \Big(\rSCS^{(0),(1,2)}_{2,g}(x_3, Y_{12}) 
	+ \rS^{(0),(1,2)}_{2}(Y_{12})\Big)
\\&\qquad\qquad\qquad
	+ \CF \Big(\rSCS^{(0),(1,2)}_{2,q}(x_1, Y_{12}) 
	+ \rSCS^{(0),(1,2)}_{2,q}(x_2, Y_{12})\Big)\Big]
\\&
	- \CA \Big[\CA \Big(\rSCS^{(0),(1,3)}_{2,g}(x_3, Y_{13}) 
	+ \rS^{(0),(1,3)}_{2}(Y_{13})\Big)
\\&\qquad
	+ \CF \Big(\rSCS^{(0),(1,3)}_{2,q}(x_1, Y_{13}) 
	+ \rSCS^{(0),(1,3)}_{2,q}(x_2, Y_{13})\Big)
\\&\qquad
	+ \CA \Big(\rSCS^{(0),(2,3)}_{2,g}(x_3, Y_{23}) 
	+ \rS^{(0),(2,3)}_{2}(Y_{23})\Big)
\\&\qquad
	+ \CF \Big(\rSCS^{(0),(2,3)}_{2,q}(x_1, Y_{23}) 
	+ \rSCS^{(0),(2,3)}_{2,q}(x_2, Y_{23})\Big)\Big]
\\&
	+ 2 \CA (2 \CF - \CA)
\\&\quad\times
	\Big(\rS^{(0),(1,2)(1,3)}_{2}(Y_{12},Y_{23},Y_{13})
	+ \rS^{(0),(1,2)(2,3)}_{2}(Y_{12},Y_{13},Y_{23})
\\&\qquad
	+ \rS^{(0),(2,3)(1,2)}_{2}(Y_{23},Y_{13},Y_{12})
	+ \rS^{(0),(1,3)(1,2)}_{2}(Y_{13},Y_{23},Y_{12})\Big)
\\&
	+ 2 \CA^2 \Big(\rS^{(0),(1,3)(2,3)}_{2}(Y_{13},Y_{12},Y_{23})
	+ \rS^{(0),(2,3)(1,3)}_{2}(Y_{23},Y_{12},Y_{13})\Big)
\\&
	+ 2(2 \CF - \CA)^2\rS^{(0),(1,2)(1,2)}_{2}(Y_{12})
\\&
	+ 2 \CA^2\Big(\rS^{(0),(1,3)(1,3)}_{2}(Y_{13}) 
	+ \rS^{(0),(2,3)(2,3)}_{2}(Y_{23})\Big)
	\bigg\}\,.
\label{eq:I23j}
\esp
\eeq
Substituting \eqnss{eq:IC20f}{eq:IS20ikik} into this expression, we obtain
\beq
\bsp
\bI_2^{(0)}(p_1,p_2,p_3;\ep) &= 
	\left[\frac{\as}{2\pi}S_\ep \left(\frac{\mu^2}{Q^2}\right)^\ep\right]^2
	\CF^2\bigg\{
		\bigg(
			2 
			+ \frac{5}{2} \colx 
			+ \frac{3}{4} \colx^2
		\bigg) \frac{1}{\ep^4} 
		+ \bigg[
			6 
			+ \frac{109}{12} \colx 
\\ &\qquad
			+ \frac{77}{24} \colx^2 
			- \frac{7}{3} \coly \Nf 
			- \frac{1}{2} \colx \coly \Nf
			-\bigg(4 + \colx - \frac{3}{2} \colx^2 \bigg) \ln y_{12}
\\ &\qquad			
			-\bigg(2 \colx + \frac{3}{2} \colx^2 \bigg) (\ln y_{13} + \ln y_{23})
		\bigg]\frac{1}{\ep^3}
		+ \Oe{-2} \bigg\} \,,
\esp
\eeq
where we used $\ln Y_{ik,Q} = \ln y_{ik} - \ln x_i - \ln x_k$. Note that the 
dependence on $y_0$ again cancels up to this order in the expansion.

The rest of the coefficients in the $\ep$-expansion are computed numerically. 
By way of illustration, we present results for the fully symmetric configuration 
of final state momenta in \tab{tab:I23jet}. In this phase-space point, the various 
invariants take the following values
\beq
x_i = \frac{2}{3}\,,
\qquad
Y_{ik,Q} = \frac{3}{4}
\qquad\mbox{and}\qquad
y_{ik} = \frac{1}{3}\,,
\eeq
where $i,k=1,2,3$ and $i\ne k$. We have again used $d_0=d'_0=3-3\ep$ and 
$\al_0=y_0=1$ during the calculation. As for the two-jet case, colour factors 
and the number of light flavours were kept symbolic. The actual values of the 
expansion coefficients are again scalar products of vectors of numbers in the 
last three columns of \tab{tab:I23jet} with the vector of coefficients which 
forms the first column. Note that the numbers presented in \tab{tab:I23jet} 
correspond to the expansion coefficients of $\bI_2^{(0)}(p_1,p_2,p_3;\ep)/\CF^2$.

\begin{table}[t]
\begin{center}
\renewcommand{\arraystretch}{2}
\begin{tabular}{|c|c|c|c|}
\hline\hline
coeff. & $1/\ep^{2}$ &  $1/\ep^{1}$ &  finite
\\
\hline\hline
\multirow{1}{*}{$1$} 
 & 17.072 $\pm$ 0.016
 & -30.766 $\pm$ 0.088
 & -253.694 $\pm$ 0.474
\\ 
\hline 
\multirow{1}{*}{$\frac{\CA}{\CF}$} 
 & 36.428 $\pm$ 0.013
 & 35.832 $\pm$ 0.068
 & -83.945 $\pm$ 0.294
\\ 
\hline 
\multirow{1}{*}{$\frac{\CA^2}{\CF^2}$} 
 & 12.580 $\pm$ 0.009
 & 3.999 $\pm$ 0.052
 & -99.299 $\pm$ 0.296
\\ 
\hline 
\multirow{1}{*}{$\frac{\Nf\TR}{\CF}$} 
 & -16.945 $\pm$ 0.003
 & -59.501 $\pm$ 0.023
 & -128.771 $\pm$ 0.144
\\ 
\hline 
\multirow{1}{*}{$\frac{\CA\Nf\TR}{\CF^2}$} 
 & -3.879 $\pm$ 0.006
 & -14.105 $\pm$ 0.052
 & -29.974 $\pm$ 0.378
\\ 
\hline\hline
\end{tabular}
\caption{
\label{tab:I23jet}
Coefficients of the Laurent expansion of 
$\displaystyle\frac{\bI_2^{(0)}(p_1,p_2,p_3;\ep)}{\CF^2}$ for $e^+e^- \to 3$ 
jet production. We used $d_0=d'_0=3-3\ep$ and $\al_0=y_0=1$.}
\end{center}
\end{table}


\section{Conclusions}
\label{sec:conclusions}

This paper finishes the calculation of the integrated doubly 
unresolved approximate cross section of the NNLO subtraction 
formalism of \Refss{Somogyi:2006da,Somogyi:2006db}, and thus 
completes the definition of the subtraction scheme.

In particular, here we computed the double soft-type contributions to the 
integrated doubly unresolved approximate cross section of \Ref{Somogyi:2006da}. 
The integrated counterterms were evaluated in terms of a set of basic double 
soft integrals, which were calculated using \MB techniques. These 
contributions represented the last missing ingredients needed to evaluate 
$\int_2\dsiga{RR}{2}_{m+2}$, as the collinear pieces of the doubly unresolved 
subtraction terms were integrated in \Ref{DelDuca:2013kw}. The final result 
can be written as the product (in colour space) of the Born cross section times 
the doubly unresolved insertion operator, $\bI_2^{(0)}$. We were able to compute 
this insertion operator analytically up to $\Oe{-2}$, while the rest of the 
expansion coefficients were evaluated numerically. 

As stressed above, the definition of the NNLO subtraction formalism of 
\Refss{Somogyi:2006da,Somogyi:2006db} is now complete, and the evaluation 
of the finite doubly virtual cross section of \eqn{eq:sigmaNNLOm} is feasible 
for electron-positron annihilation into two- and three-jets. (A few integrals 
were evaluated specifically for three-jet kinematics, so for a higher number 
of jets, some more work is required.) In particular, all integrated approximate 
cross sections appearing in \eqn{eq:sigmaNNLOm} are known analytically up to 
$\Oe{-2}$ and hence, the cancellation of poles in $\dsig{NNLO}_{m}$ may be 
checked explicitly to this order. Indeed, we have checked that the $1/\ep^4$ 
and $1/\ep^3$ poles cancel for $e^+e^-\to 2$, $3$ jets, independently of the 
value of $y_0$. The cancellation of the subleading poles must be checked 
numerically, and the details will be presented elsewhere.
As the regularised doubly real and real-virtual cross sections 
$\dsig{NNLO}_{m+2}$ and $\dsig{NNLO}_{m+1}$ of 
\eqns{eq:sigmaNNLOm+2}{eq:sigmaNNLOm+1} were previously shown to be finite 
\cite{Somogyi:2006da,Somogyi:2006db} (specifically for $m=3$), we are now in 
the position to compute the fully differential rate for electron-positron 
annihilation into two and three jets at NNLO accuracy within our framework.


\acknowledgments{
I am grateful to Vittorio Del Duca and Zolt\'an Tr\'ocs\'anyi for 
useful conversations and comments on the manuscript. 
}


\begin{appendix}


\section{Modified double soft-type subtraction terms}
\label{app:modifiedA2soft}

In previous publications, we have presented an easy modification to the NNLO 
subtraction scheme of \Refss{Somogyi:2006da,Somogyi:2006db}. The parts of 
these modifications relevant to all singly unresolved approximate cross 
sections, i.e.,~$\dsiga{RR}{1}_{m+2}$ in \eqn{eq:sigmaNNLOm+2} as well as 
$\dsiga{RV}{1}_{m+1}$ and 
$\Big(\int_1\dsiga{RR}{1}_{m+2}\Big)\strut^{{\rm A}_{\scriptscriptstyle 1}}$ 
in \eqn{eq:sigmaNNLOm+1}, were given in \Ref{Somogyi:2008fc}, while the 
parts relevant to the iterated singly unresolved approximate cross section, 
$\dsiga{RR}{12}_{m+2}$ in \eqn{eq:sigmaNNLOm+2}, were  spelled out in 
\Ref{Bolzoni:2010bt}. Finally, the modified doubly unresolved approximate 
cross section was discussed in \Ref{DelDuca:2013kw}. 

The introduction of this modification serves two purposes. First, it makes 
the integrated subtraction terms independent of $m$, the number of hard 
partons, see \Refss{Somogyi:2008fc,Bolzoni:2010bt,DelDuca:2013kw} for a 
detailed discussion. Second, it allows to constrain the subtractions to 
near the singular regions in phase space, which improves the efficiency 
of the numerical implementation.

In this paper, we only need the precise definition of the modified double 
soft-type subtraction terms. We recall from \Ref{DelDuca:2013kw} that these 
are obtained from the original subtraction terms in \Ref{Somogyi:2006da} via 
multiplication by the simple factor of $f(y_0,y_{rQ}+y_{sQ}-y_{rs},d'(m,\ep))$, 
where
\beq
f(z_0,z,p) \equiv (1-z)^{-p} \Theta(z_0-z)\,.
\eeq
We also reproduce the relevant part of Table 20 of \Ref{DelDuca:2013kw} here, 
as \tab{tab:modified-softA2}. 
\begin{table}[t]
\renewcommand\arraystretch{1.5} 
\begin{center}
\begin{tabular}{|c|c|c|}
\hline\hline
\multicolumn{3}{|c|}{Double soft-type counterterms} \\
\hline\hline
Subtraction term & Momentum mapping & Function \\
\hline
$\cS{rs}{(0,0)}$, $\cSCS{ir;s}{}\cS{rs}{(0,0)}$,
&
\multirow{3}{5.3cm}{\centering
$\mom{}_{m+2} \smap{rs} \momt{(rs)}_{m}$}
&
\multirow{3}{4.9cm}{\centering
$f(y_0, y_{rQ}+y_{sQ}-y_{rs},d'(m,\eps))$} \\
$\cC{irs}{}\cS{rs}{(0,0)}$, $\cC{ir;js}{}\cS{rs}{(0,0)}$, & & \\
$\cC{irs}{}\cSCS{ir;s}{}\cS{rs}{(0,0)}$ & & \\
\hline\hline
\end{tabular}
\caption{\label{tab:modified-softA2}
The modified double soft-type subtraction terms are obtained from 
the original counterterms (first column) by multiplication with 
$f(y_0, y_{rQ}+y_{sQ}-y_{rs},d'(m,\eps))$ (last column). Also shown 
is the momentum mapping used to define the subtraction terms (middle 
column).}
\end{center}
\end{table}
We note that the form of $d'(m,\ep)$ appearing in \tab{tab:modified-softA2} 
is actually fixed by the prescription in \Ref{Somogyi:2008fc} and the 
requirement that the modified subtraction terms still correctly regularise 
all kinematic singularities. We have
\beq
d'(m,\ep) = m(1-\ep) - d'_0\,,
\eeq
where, as in \Refss{Somogyi:2008fc,Bolzoni:2010bt,DelDuca:2013kw}, the 
parameter $d'_0$ takes the form
\beq
d'_0 = D'_0 + d'_1 \ep\,,
\label{eq:dv0def}
\eeq
with $D'_0 \ge 2$ an integer and $d'_1$ real. Throughout this paper, we use 
the specific choice $D'_0=3$ and $d'_1=-3$.
Finally, we note that in terms of $y_{\ha{r}Q}$ and $y_{sQ}$, we have
\beq
\fcut = 
	[(1-y_{\ha{r}Q})(1-y_{sQ})]^{d'_0-m(1-\ep)}
	\Theta(y_0 - y_{\ha{r}Q} - y_{sQ} + y_{\ha{r}Q} y_{sQ})\,,
\label{eq:fcut-1}
\eeq
where $y_{rs} = y_{\ha{r}s}$ and $y_{rQ} = (1-y_{sQ})y_{\ha{r}Q} + y_{\ha{r}s}$ 
have been used, see \eqn{eq:plain2hat}. 


\section{Integrating the double soft counterterms}
\label{app:RR_A2_Srs}

From \eqnss{eq:defSikjl}{eq:defSikqq} we see that the double soft counterterms 
involve the integrands
\beq
\bigg\{\calS_{ik}(r) \calS_{jl}(s)\,,\,
\calS_{ik}(r,s)\,,\,
\frac{1}{s_{rs}^2}
\bigg(\frac{s_{ir}s_{ks} + s_{kr}s_{is} - s_{ik}s_{rs}}{s_{i(rs)}s_{k(rs)}}
	-2\frac{s_{ir}s_{is}}{s_{i(rs)}^2}\bigg)\bigg\}\,,
\label{eq:Srs-integrands}
\eeq
where $\calS_{ik}(r,s)$ is given in \eqn{eq:softggnab}.

Let us proceed to identify the independent kinematic structures that we 
need to integrate. We begin with the abelian double soft gluon contribution, 
i.e.,~the first term in the list of \eqn{eq:Srs-integrands}. Recall that in 
this term, the indices $i$, $j$, $k$, $l$ are constrained only by the requirements 
that $i\ne k$ and $j\ne l$, but there is no restriction on whether or not $i$, 
$k$ is equal to $j$, $l$. Hence, we have the following situations: (i) all of $i$, 
$k$, $j$ and $l$ are distinct, (ii) only three of the four indices are distinct, 
e.g.,~$l=k$ and (iii) only two indices are distinct, e.g.,~$j=i$ and $l=k$.

For case (i) to occur, there must be at least four hard patrons in the process. 
Obviously this does not happen in the calculation of two- and three-jet 
observables, and we will not consider it in this paper. For cases (ii) and 
(iii), we have simply
\beq
\calS_{ik}(r) \calS_{jk}(s) = 
	\frac{4s_{ik}s_{jk}}{s_{ir} s_{kr} s_{js} s_{ks}}
\qquad\mbox{and}\qquad
\calS_{ik}(r) \calS_{ik}(s) = 
	\frac{4s_{ik}^2}{s_{ir} s_{kr} s_{is} s_{ks}}\,.
\label{eq:SrsggAB-MIs}
\eeq

Turning to the non-abelian part of the double soft gluon formula, the second 
term in the list of \eqn{eq:Srs-integrands}, we make two observations. First, 
the factorised phase-space measure, $[\rd p_{2;m}^{(rs)}(p_r,p_s;Q)]$, is clearly 
symmetric under the exchange of $p_r^\mu$ and $p_s^\mu$ (see \eqn{eq:dPrs2m}). 
Second, the double soft current appears in the subtraction terms summed over 
its indices, $i$ and $k$ in this case. Hence, we are free to exchange $i$ and 
$k$ in individual terms in $\calS_{ik}(r,s)$ without changing the value of the 
total integrated subtraction term, since $i$ and $k$ are merely summation indices. 
Using the freedom to make $r\leftrightarrow s$ and/or $i\leftrightarrow k$ 
replacements whenever convenient, with some algebra\footnote{We found the 
{\tt Mathematica} package {\tt \$Apart} \cite{Feng:2012iq} useful during these manipulations.} we can derive the following form of $\calS_{ik}(r,s)$,
\beq
\bsp
\calS_{ik}(r,s) &\to \frac{s_{ik}}{s_{is} s_{kr} s_{rs}} 
	- \frac{1}{2}\frac{s_{ik}^2}{s_{ir} s_{kr} s_{is} s_{ks}}
	+ (1-\ep) \bigg(\frac{1}{s_{rs}^2} - 2\frac{s_{ir} s_{kr}}{s_{i(rs)} s_{k(rs)} s_{rs}^2}\bigg)
	- 4\frac{s_{ik}}{s_{i(rs)} s_{k(rs)} s_{rs}}
\\ &
	- \frac{s_{ik}^2}{s_{i(rs)} s_{k(rs)} s_{ir} s_{kr}}
	+ 2\frac{s_{ik}}{s_{i(rs)} s_{kr} s_{rs}}\,,
\label{eq:SrsggNAB-2}
\esp
\eeq
whose integral is equal to the integral of the original expression. 

Finally, consider the expression for double soft quark-antiquark emission, the 
last term in \eqn{eq:Srs-integrands}. Making use of the freedom to exchange the 
summation indices $i$ and $k$, after some algebra we find 
\beq
\frac{1}{s_{rs}^2}
\bigg(\frac{s_{ir}s_{ks} + s_{kr}s_{is} - s_{ik}s_{rs}}{s_{i(rs)}s_{k(rs)}}
	-2\frac{s_{ir}s_{is}}{s_{i(rs)}^2}\bigg)
\to
	- 2\frac{s_{ir} s_{kr}}{s_{i(rs)} s_{k(rs)} s_{rs}^2} 
	- \frac{s_{ik}}{s_{i(rs)} s_{k(rs)} s_{rs}} 
	+ 2\frac{s_{ir}^2}{s_{i(rs)}^2 s_{rs}^2}\,. 
\label{eq:Srsqq-2}
\eeq
As in \eqn{eq:SrsggNAB-2}, the notation $\to$ above means that the two forms 
lead to the same total integrated subtraction term.

Examining \eqnss{eq:SrsggAB-MIs}{eq:Srsqq-2}, we identify the following 
independent kinematic structures\footnote{The construction of this basic 
set is not unique, and we make no claim that this set is linearly independent.} 
that we must integrate
\beq
\bsp
\bigg\{
&\frac{1}{2}\frac{s_{ik} s_{jk}}{s_{ir} s_{kr} s_{js} s_{ks}}\,,\, 
\frac{1}{2}\frac{s_{ik}^2}{s_{ir} s_{kr} s_{is} s_{ks}}\,,\,
\frac{s_{ik}}{s_{is} s_{kr} s_{rs}}\,,\, 
\frac{s_{ik}}{s_{i(rs)} s_{kr} s_{rs}}\,,\,
\\
&\frac{s_{ik}^2}{s_{i(rs)} s_{k(rs)} s_{ir} s_{kr}}\,,\,
\frac{s_{ik}}{s_{i(rs)} s_{k(rs)} s_{rs}}\,,\,
\frac{s_{ir} s_{kr}}{s_{i(rs)} s_{k(rs)} s_{rs}^2}\,,\,
\frac{1}{s_{rs}^2}\,,\,
\frac{s_{ir}^2}{s_{i(rs)}^2 s_{rs}^2}\bigg\}.
\esp
\eeq
The specific normalisations were chosen for later convenience. Finally we 
introduce the following notation,
\bal
\cI_{2\cS{}{},1}(\Yt{i}{k},\Yt{i}{j},\Yt{j}{k},\ep;y_0,d'_0) = 
	\left(\frac{16\pi^2}{S_\ep}Q^{2\ep}\right)^2
	&\int_2 [\rd p_{1;m}^{(\ha{r})}] [\rd p_{1;m+1}^{(s)}]
	\frac{1}{2}\frac{s_{ik} s_{jk}}{s_{ir} s_{kr} s_{js} s_{ks}} 
\nt\\&\times	
	\fcut\,,
\label{eq:IS1}
\\
\cI_{2\cS{}{},2}(\Yt{i}{k},\ep;y_0,d'_0) = 
	\left(\frac{16\pi^2}{S_\ep}Q^{2\ep}\right)^2
	&\int_2 [\rd p_{1;m}^{(\ha{r})}] [\rd p_{1;m+1}^{(s)}]
	\frac{1}{2}\frac{s_{ik}^2}{s_{ir} s_{kr} s_{is} s_{ks}}
\nt\\&\times	
	\fcut\,,
\label{eq:IS2}
\\
\cI_{2\cS{}{},3}(\Yt{i}{k},\ep;y_0,d'_0) = 
	\left(\frac{16\pi^2}{S_\ep}Q^{2\ep}\right)^2
	&\int_2 [\rd p_{1;m}^{(\ha{r})}] [\rd p_{1;m+1}^{(s)}]
	\frac{s_{ik}}{s_{is} s_{kr} s_{rs}}
\nt\\&\times	
	\fcut\,,
\label{eq:IS3}
\\
\cI_{2\cS{}{},4}(\Yt{i}{k},\ep;y_0,d'_0) = 
	\left(\frac{16\pi^2}{S_\ep}Q^{2\ep}\right)^2
	&\int_2 [\rd p_{1;m}^{(\ha{r})}] [\rd p_{1;m+1}^{(s)}]
	\frac{s_{ik}}{s_{i(rs)} s_{kr} s_{rs}}
\nt\\&\times	
	\fcut\,,
\label{eq:IS4}
\\
\cI_{2\cS{}{},5}(\Yt{i}{k},\ep;y_0,d'_0) = 
	\left(\frac{16\pi^2}{S_\ep}Q^{2\ep}\right)^2
	&\int_2 [\rd p_{1;m}^{(\ha{r})}] [\rd p_{1;m+1}^{(s)}]
	\frac{s_{ik}^2}{s_{i(rs)} s_{k(rs)} s_{ir} s_{kr}}
\nt\\&\times	
	\fcut\,,
\label{eq:IS5}
\\
\cI_{2\cS{}{},6}(\Yt{i}{k},\ep;y_0,d'_0) = 
	\left(\frac{16\pi^2}{S_\ep}Q^{2\ep}\right)^2
	&\int_2 [\rd p_{1;m}^{(\ha{r})}] [\rd p_{1;m+1}^{(s)}]
	\frac{s_{ik}}{s_{i(rs)} s_{k(rs)} s_{rs}}
\nt\\&\times	
	\fcut\,,
\label{eq:IS6}
\\
\cI_{2\cS{}{},7}(\Yt{i}{k},\ep;y_0,d'_0) = 
	\left(\frac{16\pi^2}{S_\ep}Q^{2\ep}\right)^2
	&\int_2 [\rd p_{1;m}^{(\ha{r})}] [\rd p_{1;m+1}^{(s)}]
	\frac{s_{ir} s_{kr}}{s_{i(rs)} s_{k(rs)} s_{rs}^2}
\nt\\&\times	
	\fcut\,,
\label{eq:IS7}
\\
\cI_{2\cS{}{},8}(\ep;y_0,d'_0) = 
	\left(\frac{16\pi^2}{S_\ep}Q^{2\ep}\right)^2
	&\int_2 [\rd p_{1;m}^{(\ha{r})}] [\rd p_{1;m+1}^{(s)}]
	\frac{1}{s_{rs}^2}
\nt\\&\times	
	\fcut\,,
\label{eq:IS8}
\\
\cI_{2\cS{}{},9}(\ep;y_0,d'_0) = 
	\left(\frac{16\pi^2}{S_\ep}Q^{2\ep}\right)^2
	&\int_2 [\rd p_{1;m}^{(\ha{r})}] [\rd p_{1;m+1}^{(s)}]
	\frac{s_{ir}^2}{s_{i(rs)}^2 s_{rs}^2}
\nt\\&\times	
	\fcut\,.
\label{eq:IS9}
\eal
In terms of these basic integrals, the double soft integrated counterterms 
are expressed as in \eqnss{eq:ISrs0ikjkMIexp}{eq:ISrs0qqikMIexp}.


\section{Integrating the triple collinear -- double soft counterterms}
\label{app:RR_A2_CirsSrs}

From \eqnss{eq:defC3Sab}{eq:defC3Sqq}, we find that the triple collinear -- 
double soft counterterms involve the following integrands
\beq
\bsp
\bigg\{
&\frac{4 \tzz{i}{rs}^2}{s_{ir} s_{is} \tzz{r}{is} \tzz{s}{ir}}\,,\,
\frac{(1-\ep)}{s_{i(rs)}s_{rs}} 
	\frac{(s_{ir} \tzz{s}{ir} - s_{is}\tzz{r}{is})^2}
	{s_{i(rs)}s_{rs}(\tzz{r}{is} + \tzz{s}{ir})^2}
	-\frac{\tzz{i}{rs}}{s_{i(rs)}s_{rs}}
	\bigg(\frac{4}{\tzz{r}{is} + \tzz{s}{ir}} - \frac{1}{\tzz{r}{is}}\bigg)
\\&
	-\frac{1}{s_{i(rs)}s_{ir}}
	\frac{2\tzz{i}{rs}^2}{\tzz{r}{is}(\tzz{r}{is} + \tzz{s}{ir})}
	-\frac{\tzz{i}{rs}^2}{s_{i(rs)}s_{is}}
	\frac{1}{\tzz{r}{is}(\tzz{r}{is} + \tzz{s}{ir})}
	+\frac{\tzz{i}{rs}}{s_{ir}s_{rs}}
	\bigg(\frac{1}{\tzz{s}{ir}} + \frac{1}{\tzz{r}{is} + \tzz{s}{ir}}\bigg)\,,\,
\\&
\frac{2}{s_{i(rs)}s_{rs}} 
	\bigg(\frac{\tzz{i}{rs}}{\tzz{r}{is} + \tzz{s}{ir}}
	-\frac{(s_{ir} \tzz{s}{ir} - s_{is} \tzz{r}{is})^2}
	{s_{i(rs)}s_{rs}(\tzz{r}{is} + \tzz{s}{ir})^2}\bigg)\bigg\}\,.
\esp
\label{eq:CirsSrs-integrands}
\eeq
Note that we have not written the $r\leftrightarrow s$ part of the non-abelian 
triple collinear -- double soft gluon counterterm (see \eqn{eq:defC3Snab}) since 
by the $r\leftrightarrow s$ symmetry of the factorised phase-space measure, its 
integral is equal to the integral of the displayed term.

Now, we would like to identify a basic set of kinematic structures, which 
we must integrate. At first glance, it would seem like we ought to use the 
identity $\tzz{i}{rs}+\tzz{r}{is}+\tzz{s}{ir}=1$ to eliminate one of the $z$'s. 
However it turns out that it is more convenient to simply keep the integrand 
essentially in its original form, as given in  \eqn{eq:CirsSrs-integrands}, for 
reasons we shall explain shortly. Then we find that up to an $r\leftrightarrow s$ 
exchange, there are nine basic structures to integrate. These are:
\beq
\bsp
\bigg\{
&\frac{1}{s_{ir} s_{is}} 
	\frac{\tzz{i}{rs}^2}{\tzz{r}{is} \tzz{s}{ir}}\,,\,
\frac{s_{ir}^2}{s_{i(rs)}^2 s_{rs}^2} 
	\frac{\tzz{s}{ir}^2}{(\tzz{r}{is} + \tzz{s}{ir})^2}\,,\,
\frac{s_{ir} s_{is}}{s_{i(rs)}^2 s_{rs}^2} 
	\frac{\tzz{r}{is} \tzz{s}{ir}}{(\tzz{r}{is} + \tzz{s}{ir})^2}\,,\,
\frac{1}{s_{i(rs)} s_{rs}} 
	\frac{\tzz{i}{rs}}{(\tzz{r}{is} + \tzz{s}{ir})}\,,\,
\\
&\frac{1}{s_{i(rs)} s_{rs}} 
	\frac{\tzz{i}{rs}}{\tzz{r}{is}}\,,\,
\frac{1}{s_{i(rs)} s_{ir}} 
	\frac{\tzz{i}{rs}^2}{\tzz{r}{is}(\tzz{r}{is} + \tzz{s}{ir})}\,,\,
\frac{1}{s_{i(rs)} s_{is}} 
	\frac{\tzz{i}{rs}^2}{\tzz{r}{is}(\tzz{r}{is} + \tzz{s}{ir})}\,,\,
\frac{1}{s_{ir} s_{rs}} 
	\frac{\tzz{i}{rs}}{\tzz{r}{is}}\,,\,
\\
&\frac{1}{s_{ir} s_{rs}} 
	\frac{\tzz{i}{rs}}{(\tzz{r}{is} + \tzz{s}{ir})}\bigg\}\,.
\esp
\eeq
The reason the above integrands are convenient is as follows: clearly the results 
of all these integrals could only depend on $x_{\wti{i}}=2\ti{p}_i\cdot Q/Q^2$. 
However, all nine of the above integrands are {\em degree zero homogenous} functions 
of $p_i$. Hence, they are actually {\em independent} of $x_{\wti{i}}$, and thus 
are free of any kinematic dependence. This nice property is clearly lost if we 
decompose e.g.,~$\tzz{i}{rs}$ in the numerators as $\tzz{i}{rs} = 1 - \tzz{r}{is} 
- \tzz{s}{ir}$ and perform partial fractioning. Then we introduce the notation
\bal
\cI_{2\cS{}{},10}(\ep;y_0,d'_0) &= 
	\left(\frac{16\pi^2}{S_\ep}Q^{2\ep}\right)^2
	\int_2 [\rd p_{1;m}^{(\ha{r})}] [\rd p_{1;m+1}^{(s)}]
	\frac{1}{s_{ir} s_{is}} 
	\frac{\tzz{i}{rs}^2}{\tzz{r}{is} \tzz{s}{ir}}
\nt\\&\times	
	\fcut\,,
\label{eq:IS10}
\\
\cI_{2\cS{}{},11}(\ep;y_0,d'_0) &= 
	\left(\frac{16\pi^2}{S_\ep}Q^{2\ep}\right)^2
	\int_2 [\rd p_{1;m}^{(\ha{r})}] [\rd p_{1;m+1}^{(s)}]
	\frac{s_{ir}^2}{s_{i(rs)}^2 s_{rs}^2} 
	\frac{\tzz{s}{ir}^2}{(\tzz{r}{is} + \tzz{s}{ir})^2}
\nt\\&\times	
	\fcut\,,
\label{eq:IS11}
\\
\cI_{2\cS{}{},12}(\ep;y_0,d'_0) &= 
	\left(\frac{16\pi^2}{S_\ep}Q^{2\ep}\right)^2
	\int_2 [\rd p_{1;m}^{(\ha{r})}] [\rd p_{1;m+1}^{(s)}]
	\frac{s_{ir} s_{is}}{s_{i(rs)}^2 s_{rs}^2} 
	\frac{\tzz{r}{is} \tzz{s}{ir}}{(\tzz{r}{is} + \tzz{s}{ir})^2}
\nt\\&\times	
	\fcut\,,
\label{eq:IS12}
\\
\cI_{2\cS{}{},13}(\ep;y_0,d'_0) &= 
	\left(\frac{16\pi^2}{S_\ep}Q^{2\ep}\right)^2
	\int_2 [\rd p_{1;m}^{(\ha{r})}] [\rd p_{1;m+1}^{(s)}]
	\frac{1}{s_{i(rs)} s_{rs}} 
	\frac{\tzz{i}{rs}}{(\tzz{r}{is} + \tzz{s}{ir})}
\nt\\&\times	
	\fcut\,,
\label{eq:IS13}
\\
\cI_{2\cS{}{},14}(\ep;y_0,d'_0) &= 
	\left(\frac{16\pi^2}{S_\ep}Q^{2\ep}\right)^2
	\int_2 [\rd p_{1;m}^{(\ha{r})}] [\rd p_{1;m+1}^{(s)}]
	\frac{1}{s_{i(rs)} s_{rs}} 
	\frac{\tzz{i}{rs}}{\tzz{r}{is}}
\nt\\&\times	
	\fcut\,,
\label{eq:IS14}
\\
\cI_{2\cS{}{},15}(\ep;y_0,d'_0) &= 
	\left(\frac{16\pi^2}{S_\ep}Q^{2\ep}\right)^2
	\int_2 [\rd p_{1;m}^{(\ha{r})}] [\rd p_{1;m+1}^{(s)}]
	\frac{1}{s_{i(rs)} s_{ir}} 
	\frac{\tzz{i}{rs}^2}{\tzz{r}{is}(\tzz{r}{is} + \tzz{s}{ir})}
\nt\\&\times	
	\fcut\,,
\label{eq:IS15}
\\
\cI_{2\cS{}{},16}(\ep;y_0,d'_0) &= 
	\left(\frac{16\pi^2}{S_\ep}Q^{2\ep}\right)^2
	\int_2 [\rd p_{1;m}^{(\ha{r})}] [\rd p_{1;m+1}^{(s)}]
	\frac{1}{s_{i(rs)} s_{is}} 
	\frac{\tzz{i}{rs}^2}{\tzz{r}{is}(\tzz{r}{is} + \tzz{s}{ir})}
\nt\\&\times	
	\fcut\,,
\label{eq:IS16}
\\
\cI_{2\cS{}{},17}(\ep;y_0,d'_0) &= 
	\left(\frac{16\pi^2}{S_\ep}Q^{2\ep}\right)^2
	\int_2 [\rd p_{1;m}^{(\ha{r})}] [\rd p_{1;m+1}^{(s)}]
	\frac{1}{s_{ir} s_{rs}} 
	\frac{\tzz{i}{rs}}{\tzz{r}{is}}
\nt\\&\times	
	\fcut\,,
\label{eq:IS17}
\\
\cI_{2\cS{}{},18}(\ep;y_0,d'_0) &= 
	\left(\frac{16\pi^2}{S_\ep}Q^{2\ep}\right)^2
	\int_2 [\rd p_{1;m}^{(\ha{r})}] [\rd p_{1;m+1}^{(s)}]
	\frac{1}{s_{ir} s_{rs}} 
	\frac{\tzz{i}{rs}}{(\tzz{r}{is} + \tzz{s}{ir})}
\nt\\&\times	
	\fcut\,.
\label{eq:IS18}
\eal
The triple collinear -- double soft integrated counterterms are 
expressed with these basic integrals as in \eqnss{eq:ICirsSrs0ggABMIexp}
{eq:ICirsSrs0qqMIexp}.


\section{Integrating the double collinear -- double soft counterterm}
\label{app:RR_A2_CirjsSrs}

Recall from \eqn{eq:defC2S} that the integrated double collinear -- double 
soft counterterm involves the integrand
\beq
\bigg\{
	\frac{1}{s_{ir}} \frac{2\tzz{i}{r}}{\tzz{r}{i}}
	\frac{1}{s_{js}} \frac{2\tzz{j}{s}}{\tzz{s}{j}}
\bigg\}\,.
\label{eq:CirjsSrs-integrands}
\eeq

We again keep the integrand in its original form, without making use of 
the identities $\tzz{i}{r}+\tzz{r}{i}=1$ and $\tzz{j}{s}+\tzz{s}{j}=1$. 
The reason is clear: since the expression in \eqn{eq:CirjsSrs-integrands} 
is degree zero homogenous in both $p_i$ and $p_j$, there can be no direct 
dependence on $x_{\wti{i}}$ or $x_{\wti{j}}$. However, since the phase-space 
measure is not exactly factorised in $p_r^\mu$ and $p_s^\mu$ (it is exactly 
factorised in $\ha{p}_r^\mu$ and $p_s^\mu$), the integral still depends on 
$\Yt{i}{j}$. Introducing the notation
\bal
\cI_{2\cS{}{},19}(\Yt{i}{j},\ep;y_0,d'_0) &= 
	\left(\frac{16\pi^2}{S_\ep}Q^{2\ep}\right)^2
	\int_2 [\rd p_{1;m}^{(\ha{r})}] [\rd p_{1;m+1}^{(s)}]
	\frac{1}{s_{ir} s_{js}} 
	\frac{\tzz{i}{r} \tzz{j}{s}}{\tzz{r}{i} \tzz{s}{j}}
\nt\\&\times	
	\fcut\,,
\label{eq:IS19}
\eal
(note the normalisation) we immediately find \eqn{eq:ICirjsSrs0MIexp}.


\section{Integrating the soft collinear -- double soft counterterm}
\label{app:RR_A2_CSirsSrs}

Form \eqn{eq:defCSSjl} we see that the integrated soft collinear -- double 
soft counterterm involves the integrand
\beq
\bigg\{\frac{1}{s_{ir}} \frac{\tzz{i}{r}}{\tzz{r}{i}}\calS_{jl}(s)\Bigg\}\,,
\label{eq:CSirsSrs-integrand}
\eeq
where $j=(ir)$ or $l=(ir)$ is also allowed. (Recall that at the level of 
kinematics, e.g.,~$j=(ir)$ means that the momentum $j$ entering the eikonal 
factor is $p_j^\mu=p_i^\mu + p_r^\mu$.)

Using the explicit expression of the eikonal factor, \eqn{eq:Sikr-eikonal}, 
we find that \eqn{eq:CSirsSrs-integrand} evaluates as
\beq
\bigg\{
\frac{1}{s_{ir}} 
	\frac{\tzz{i}{r}}{\tzz{r}{i}} \frac{s_{jl}}{s_{js} s_{ls}}\,,\,
\frac{1}{s_{ir}} 
	\frac{\tzz{i}{r}}{\tzz{r}{i}} \frac{s_{(ir)l}}{s_{(ir)s} s_{ls}}\bigg\}\,,
\eeq
where the first expression corresponds to the case when both $j$ and $l$ are 
distinct from $(ir)$, while the second corresponds to $j=(ir)$. We introduce the 
notation
\bal
\cI_{2\cS{}{},20}(\Yt{j}{l},\ep;y_0,d'_0) &= 
	\left(\frac{16\pi^2}{S_\ep}Q^{2\ep}\right)^2
	\int_2 [\rd p_{1;m}^{(\ha{r})}] [\rd p_{1;m+1}^{(s)}]
	\frac{1}{s_{ir}} 
	\frac{\tzz{i}{r}}{\tzz{r}{i}} \frac{s_{jl}}{s_{js} s_{ls}}
\nt\\&\times	
	\fcut\,,
\label{eq:IS20}
\\
\cI_{2\cS{}{},21}(x_{\wti{i}},\Yt{i}{l},\ep;y_0,d'_0) &= 
	\left(\frac{16\pi^2}{S_\ep}Q^{2\ep}\right)^2
	\int_2 [\rd p_{1;m}^{(\ha{r})}] [\rd p_{1;m+1}^{(s)}]
	\frac{1}{s_{ir}} 
	\frac{\tzz{i}{r}}{\tzz{r}{i}} \frac{s_{(ir)l}}{s_{(ir)s} s_{ls}}
\nt\\&\times	
	\fcut\,,
\label{eq:IS21}
\eal
and obtain the soft collinear -- double soft integrated counterterm as in
\eqns{eq:ICSirsSrs0jlMIexp}{eq:ICSirsSrs0irlMIexp}, for $j, l \ne (ir)$ 
and $j=(ir)$ respectively.


\section{Integrating the triple collinear -- soft collinear -- double soft 
counterterm}
\label{app:RR_A2_CirsCSirsSrs}

Recall from \eqn{eq:defCCSSgg} that the integrated triple collinear -- soft 
collinear -- double soft counterterm involves the integrand
\beq
\bigg\{	\frac{4\tzz{i}{rs}(\tzz{i}{rs} + \tzz{r}{is})}
	{s_{ir}s_{(ir)s}\tzz{r}{is}\tzz{s}{ir}} \bigg\}\,.
\eeq

In this case, we will simply split the sum in the numerator and use as our 
independent integrals the following:
\beq
\bigg\{
\frac{1}{s_{(ir)s} s_{ir}} 
	\frac{\tzz{i}{rs}^2}{\tzz{r}{is}^2 \tzz{s}{ir}^2}\,,\,
\frac{1}{s_{(ir)s} s_{ir}} 
	\frac{\tzz{i}{rs}}{\tzz{s}{ir}}
\bigg\}\,.
\eeq
With the notation
\bal
\cI_{2\cS{}{},22}(x_{\wti{i}},\ep;y_0,d'_0) &= 
	\left(\frac{16\pi^2}{S_\ep}Q^{2\ep}\right)^2
	\int_2 [\rd p_{1;m}^{(\ha{r})}] [\rd p_{1;m+1}^{(s)}]
	\frac{1}{s_{(ir)s} s_{ir}} 
	\frac{\tzz{i}{rs}^2}{\tzz{r}{is}^2 \tzz{s}{ir}^2}
\nt\\&\times	
	\fcut\,,
\label{eq:IS22}
\\
\cI_{2\cS{}{},23}(x_{\wti{i}},\ep;y_0,d'_0) &= 
	\left(\frac{16\pi^2}{S_\ep}Q^{2\ep}\right)^2
	\int_2 [\rd p_{1;m}^{(\ha{r})}] [\rd p_{1;m+1}^{(s)}]
	\frac{1}{s_{(ir)s} s_{ir}} 
	\frac{\tzz{i}{rs}}{\tzz{s}{ir}}
\nt\\&\times	
	\fcut\,,
\label{eq:IS23}
\eal
(note the normalisation) we immediately find \eqn{eq:ICirsCSirsSrs0MIexp}.


\section{Collinear-type doubly unresolved counterterms to $\Oe{-2}$}
\label{app:collICTs}

In this appendix we recall from \Ref{DelDuca:2013kw} the analytic formulae 
for the flavour-summed collinear-type functions in \eqn{eq:cali}. We have
 
\begin{enumerate} 
\item Triple collinear: 
\bal 
\Big(\IcC{irs}{(0)}\Big)_{f_i}(x_i) &=  
	{1 \over 2}\bigg(1 + {\CA \over 2 C_{f_i}}\bigg)  
	\bigg[ 
	{1 \over \ep^4} - {1 \over \ep^3} \bigg(4 \ln x_i - \gam{f_i}(\Nf)
	- \gam{f_i}\bigg(\frac{\CF}{C_{f_i}}\Nf\bigg) \bigg) \bigg] 
\nt\\ &
	+ {1 \over \ep^3} {\CA \over 4 C_{f_i}} 
	\gam{g}\bigg(\frac{\CF}{C_{f_i}}\Nf\bigg) 
	+ \Oe{-2} \,. 
\eal 

\item Triple collinear -- soft collinear: 
\bal 
\Big(\IcC{irs}{}\IcSCS{ir;s}{(0)}\Big)_{f_i}(x_i) &= 
	{2 \over 3} \bigg[{1 \over \ep^4} - {2 \over \ep^3}  
	\bigg(\ln x_i + \Sigma(y_0, D'_0 - 1)\bigg)\bigg]  
	+ {1 \over 2\ep^3} \gam{f_i}(\Nf) 
\nt\\ &
	+ \Oe{-2} \,. 
\eal 

\item Double collinear: 
\bal 
\Big(\IcC{ir;js}{(0)}\Big)_{f_if_j}(x_i,x_j) &= 
	{1 \over 2 \ep^4}  
	- {1 \over 2 \ep^3} \bigg[ 2 ( \ln x_i + \ln x_j ) 
	- \gam{f_i}(\Nf) - \gam{f_j}(\Nf) \bigg] 
\nt\\ &
	+ \Oe{-2} \,. 
\eal 

\item Double collinear -- soft collinear: 
\bal 
\Big(\IcC{ir;js}{}\IcSCS{ir;s}{(0)}\Big)_{f_i f}(x_i,Y_{ij;Q}) &= 
	{1 \over \ep^4}  
	- {2 \over \ep^3}\bigg(\ln x_i + \Sigma(y_0,D'_0-1)\bigg) 
	+ {1 \over \ep^3} \gam{f_i}(\Nf) 
\nt\\ &
	+ \Oe{-2} \,. 
\eal 

\item Soft collinear: 
\bal 
\Big(\IcSCS{ir;s}{(0)}\Big)_{f_i}^{(j,l)}(x_i,Y_{jl;Q}) &= 
	- {1 \over \ep^4}  
	+ {2 \over \ep^3}\bigg(\ln x_i + \Sigma(y_0,D'_0-1)\bigg) 
	+ {1 \over \ep^3} \big(\ln Y_{jl;Q} - \gam{f_i}(\Nf)\big) 
\nt\\ &
	+ \Oe{-2} \,, 
\eal
for $j\ne i$, and 
\bal 
\Big(\IcSCS{ir;s}{(0)}\Big)_{f_i}^{(i,l)}(x_i,Y_{il;Q}) &= 
	{5 \over 6} \bigg[- {1 \over \ep^4}  
	+ {2 \over \ep^3} \bigg(\ln x_i  + \Sigma(y_0,D'_0-1)\bigg)\bigg] 
\nt\\ &
	+ {1 \over \ep^3} \bigg( \ln Y_{il;Q}- {3 \over 4} \gam{f_i}(\Nf)\bigg) 
	+ \Oe{-2} \,. 
\eal 
for $j=i$.
\end{enumerate} 

\end{appendix}



\providecommand{\href}[2]{#2}\begingroup\raggedright\endgroup


\end{document}